\documentclass[a4paper,11pt]{article}
\pdfoutput=1 

\usepackage{jheppub} 
\usepackage{framed}
\usepackage{enumerate}
\usepackage{amsmath,amssymb}
\usepackage{float}
\usepackage{xcolor}
\usepackage[T1]{fontenc} 
\usepackage{epigraph}

\usepackage{amssymb}
\usepackage{amsmath}
\usepackage{graphicx}
\usepackage{subfig}
\usepackage{cancel}
\usepackage{dsfont}

\newcommand{\la}[1]{\label{#1}}
\newcommand{\eq}[1]{\eqref{#1}}

\def\[{\left[}
\def\]{\right]}
\def\({\left(}
\def\){\right)}
\def\d{\partial}
\newcommand{\beq}{\begin{equation}}
\newcommand{\eeq}{\end{equation}}
\newcommand\beqa{\begin{eqnarray}}
\newcommand\eeqa{\end{eqnarray}}
\newcommand{\nn}{\nonumber}

\newcommand{\oneD}{{1 {\rm D}}}

\definecolor{color1}{rgb}{0.471412, 0.108766, 0.527016}
\definecolor{color2}{rgb}{0.246296, 0.31595666666666666, 0.80044}
\definecolor{color3}{rgb}{0.324106, 0.6089696666666666, 0.7083413333333334}
\definecolor{color4}{rgb}{0.513417, 0.72992, 0.440682}
\definecolor{color5}{rgb}{0.764712, 0.7283023333333333, 0.27360833333333334}
\definecolor{color6}{rgb}{0.901627, 0.5398719999999999, 0.208366}
\definecolor{color7}{rgb}{0.857359, 0.131106, 0.132128}

\title{\boldmath Integrated correlators from integrability: Maldacena-Wilson line in ${\cal N}=4$ SYM

}

\author{Andrea Cavagli\`a$^a$}
\author{Nikolay Gromov$^{b}$}
\author{Julius Julius$^{b}$}
\author{Michelangelo Preti$^b$}%
\affiliation{%
  $^a$ Dipartimento di Fisica, Universit\`a di Torino and INFN - Sezione di Torino\\
  Via P. Giuria 1, 10125 Torino, Italy 
 \\
 $^b$ Department of Mathematics, King's College London\\
 Strand WC2R 2LS, London, UK 
}%
 \emailAdd{andrea.cavaglia@unito.it}
 \emailAdd{nikolay.gromov@kcl.ac.uk}
 \emailAdd{julius.julius@kcl.ac.uk}
 \emailAdd{michelangelo.preti@kcl.ac.uk}

\abstract{
We present a  systematic method for the derivation of a relation which connects the correlation 
function of operators on the straight Maldacena-Wilson line with the integrability data for the cusp anomalous dimension. As we show, the derivation requires very careful treatment of the UV divergences. Our method opens a way to derive infinitely many  constraints on integrals of  multi-point correlation functions,  relating them with the integrability data for the generalised cusp anomalous dimension governed by the Quantum Spectral Curve.
Such constraints have been shown recently to be very powerful in  combination with the numerical conformal bootstrap, leading to very narrow non-perturbative bounds on conformal data beyond the spectrum.
}

\begin{document} 
\maketitle
\flushbottom
\newpage
\section{Introduction}
\label{sec:intro}

Recently, the combination of the exact techniques of integrability and the conformal bootstrap has proven to be very powerful for the non-perturbative study of beyond-the-spectrum observables in higher-dimensional interacting conformal field theories (CFT) such as ${\cal N}=4$ SYM~\cite{Cavaglia:2021bnz,Cavaglia:2022qpg,Caron-Huot:2022sdy}.
The main idea of this program, called \emph{Bootstrability}, is to inject non-perturbative spectral information -- obtained using a powerful integrability-based method called Quantum Spectral Curve (QSC)~\cite{Gromov:2013pga,Gromov:2014caa}
-- into the crossing equations of the CFT of interest. Even without knowing the spectrum, the  methods of the numerical conformal bootstrap (NCB)~\cite{Rattazzi:2008pe,El-Showk:2012cjh,Poland:2018epd,Chester:2019wfx}  allow one to obtain bounds on various observables (mainly the spectrum but also structure constants), see e.g. for applications to  AdS/CFT  \cite{Agmon:2017xes,Beem:2013qxa,Beem:2016wfs,Chester:2021aun}. Especially at strong coupling, those bounds become very narrow allowing for analytic studies too~\cite{Aharony:2016dwx,Alday:2017xua,Alday:2022uxp,Alday:2022xwz}. In  combination with the methods of integrability,  one can focus more sharply on the structure constants, and obtain extremely narrow bounds on the  latter~\cite{Cavaglia:2021bnz,Cavaglia:2022qpg}, as shown on figure~\ref{Cplot}, with the leading OPE coefficient determined with the error as small as $10^{-6}$ at the 't Hooft coupling $\lambda \sim 25$, and the bound rapidly shrinking at stronger and weak coupling.

\begin{figure}
    \centering
    \includegraphics[scale={1.4}]{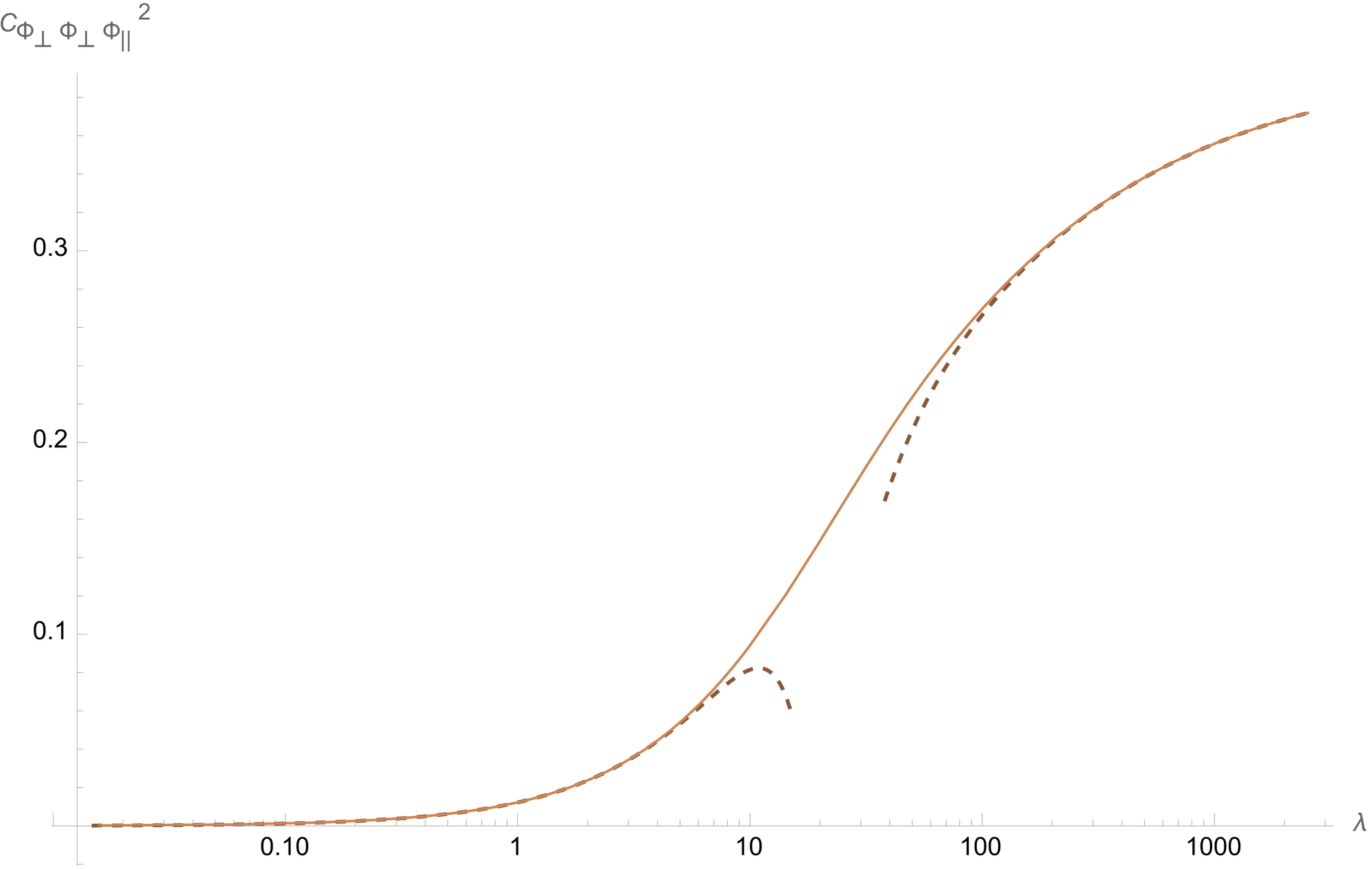}
    \caption{\la{Cplot}High precision non-perturbative evaluation of the structure constant of two protected one non-protected operators $C^2_{\Phi_{\perp}\Phi_{\perp}\Phi_{||}}$ in the defect CFT living on the Maldacena-Wilson line, as obtained by \cite{Cavaglia:2022qpg} (solid line). Dashed lines represent weak and strong coupling analytic results of \cite{Kiryu:2018phb,Cavaglia:2022qpg,Ferrero:2021bsb}}
    \label{fig:my_label}
\end{figure}

In particular, in~\cite{Cavaglia:2022qpg} it was realised that, in addition to the spectrum of the CFT itself, one can inject even more spectral data  coming from integrable deformations of the setup. In the setup considered here, which we review in detail below, the relevant integrable deformation is that of a supersymmetric straight Maldacena-Wilson line (MWL)  deformed by forming a cusp. The extra spectral data pertaining to the anomalous dimension of the cusp impose  additional  constraints on the conformal data of the defect CFT, on top of the crossing equations. 
These constraints  manifest themselves as extra relations on integrated multi-point functions. Including those in the bootstrap dramatically improves the bounds on the OPE coefficients (shrinking them by $4-5$ orders of magnitude for the leading one,  as compared to when the spectrum alone is included!~\cite{Cavaglia:2022qpg}).  Similar relations on the integrated four-point function of four local operators in ${\cal  N} = 4$ SYM have recently been obtained using localisation, rather than integrability, in \cite{Binder:2019jwn,Paul:2022piq} (see also~\cite{Dorigoni:2021bvj,Dorigoni:2021guq,Dorigoni:2022zcr}), and have been shown to also greatly help improving the bounds given by the  NCB~\cite{Chester:2021aun}. 
The advantage of the integrability-generated constraints is that, in principle, they allow one to also constrain correlators with non-BPS external operators. Moreover, in general they are complementary to the information one can obtain from  localisation. Hence, there is a good chance that the combination of the two approaches could reinforce results further, even for the case of local operators.

In~\cite{Cavaglia:2022qpg}, we found and already used two such constraints for the correlators of 4 point correlators on top of a MWL. 
Recently, in~\cite{Drukker:2022pxk}, a specific linear combination of these constraints was derived using  general geometric arguments about the structure of the conformal manifold. 
In this paper, we present a very different argument,  which allowed us to derive the remaining independent  constraint. 
As we will show, the derivation requires very careful treatment of the UV divergences and contains many technically involved steps  (mostly detailed in the appendices of this paper), but should allow for further generalisations. In particular, one should be able to obtain constraints on 6-point functions, of the type studied recently at weak coupling in \cite{Barrat:2021tpn,Barrat:2022eim}.

This paper is organised as follows. In section~\ref{sec:setup} we describe the setup of the one-dimensional CFT together with its deformations as well as the integral constraints found in~\cite{Cavaglia:2022qpg} in a notation that we will use for our derivation.  In section~\ref{sec:CPT} we explain our main strategy and introduce the tools of conformal perturbation theory, as well as describing our regularisation scheme. 
Finally, in section~\ref{sec:derive} we present our main result, the derivation of an integrated correlator related to the Curvature function known from integrability. Technical details are collected in the appendices.

\section{Setup}\label{sec:setup}

In this section we describe in detail our setup. After the definition of supersymmetric Wilson line, we describe the properties of the one-dimensional CFT which lives on top of it.
We introduce also the deformations of this CFT and their relation to insertions of local operators and the two crucial quantities -- the Bremsstrahlung and Curvature functions. Finally, we review the integrated correlators found initially in \cite{Cavaglia:2022qpg}.
 
\subsection{The $1/2$ BPS Maldacena-Wilson line}\label{sec:line}

We consider the one dimensional defect created by the supersymmetric Maldacena-Wilson loop (MWL) \cite{Maldacena:1998im} in 4D $\mathcal{N}=4$ SYM, defined as follows
\begin{align}
\label{eqn:MWLdefine}
    {\cal W}_{\mathcal{C}} =\frac{1}{N} \operatorname{Tr} \operatorname{P}\exp\int_{\mathcal{C}}dt\,\left(i\, A_{\mu}\dot{x}(t)^\mu + |\dot{x}(t)|\;n\cdot\Phi\right)\;,
\end{align}
where the contour $\mathcal{C}$ is parametrised by $x^\mu(t)$. The trace $\operatorname{Tr}$ is taken in the fundamental representation and $\operatorname{P}$ stands for the path-ordering. The coupling to the scalars $\vec{n}$ is a six-dimensional vector with unit norm.

In this paper we consider the case when ${\cal C}$ is a straight line (or a circle). In this case the defect preserves half of the supercharges of the full theory. We
also fix $\vec{n}=\{0,0,0,0,0,1\}$
and introduce the notation $\Phi_{||}=\vec n\cdot \Phi =  \Phi^6$, while the remaining $5$ scalars we denote as $\Phi_{\perp}^{M}$ with $M = 1,\cdots, 5$ (with capital Latin indexes). The full  ${\rm SO}(6)$ $R$-symmetry of the parent theory is thus broken to ${\rm SO}(5)$ by the MWL. Similarly, the full 
superconformal symmetry ${\rm PSU}(2,2|4)$ of $\mathcal{N}=4$ SYM is broken in the presence of the defect to ${\rm OSp}(4^*|4)$, where the bosonic subgroups are the ${\rm SO}(3)$ rotations about the loop and  the 1D conformal group ${\rm SO}(1,2)$ preserving the line. 

Despite the MWLs lying on straight lines and circles preserve the same symmetries and are related by a conformal transformation, their expectation values are different due to a subtle anomaly~ \cite{Drukker:2000rr}. More precisely for the straight line, in some natural conventions, $\langle \mathcal{W}_{\text{line}} \rangle = 1$~\cite{Drukker:1999zq,Erickson:2000af,Zarembo:2002an}, at the same time the vev of the circular MWL depends on the coupling constant and at large $N$ that is given by \cite{Erickson:2000af,Drukker:2000rr,Pestun:2009nn}
\beq\label{WLcircle}
\langle \mathcal{W}_{\text{circle}} \rangle=
\frac{1}{2\pi g}I_1(4\pi g)\,,
\eeq
where $g$ is the 't Hooft coupling $g=\tfrac{\sqrt{\lambda}}{4\pi}$ and $I_n(z)$ is the modified Bessel function of the first kind. Apart from this curiosity the circle and the straight line are pretty much the same, but for technical reasons we will be working on the circle most of the time in this paper.

\subsection{The defect CFT living on the $1/2$ BPS MWL}\label{sec:1dcft}

The straight MWL \eqref{eqn:MWLdefine}  can be interpreted as a superconformal one-dimensional defect. The defect theory on top of this defect possesses all the standard properties of a conformal field theory. Its operators are realised by inserting $\mathcal{N}$=4 SYM fields along the Wilson line. Operators arrange in (super)multiplets identified by representations of the symmetry unbroken by the  defect, 2- and 3-point functions kinematics is constrained and higher point functions can be  constructed by the Operator Product Expansion (OPE). In the following we review these general facts.

\paragraph{The states of the 1D defect CFT.} Operators are organised in superconformal multiplets labelled by four quantum numbers according to the unitary representations of the unbroken ${\rm OSp}(4^*|4)$ symmetry \cite{Gunaydin:1990ag,Liendo:2016ymz}. The quantum numbers can be represented in the form $\{\Delta, \[a,b\], s\}$ where $\Delta$ is the scaling dimension, $\[a,b\]$ are the Dynkin labels associated with the $R$-symmetry and $s$ is the spin associated with rotations about the line. At generic values of finite coupling, the 1D CFT admits two classes of supermultiplets. 
The simplest ones are the $1/2$-BPS short multiplets denoted as ${\cal B}_{k}$, whose superconformal primaries have scaling dimensions protected by supersymmetry with quantum numbers $\{n,\[0,n\],0\}$ with $n\in\mathbb{Z}$. In addition to these, the defect theory admits long multiplets ${\cal L}^{\Delta}_{s,\[a,b\]}$ which in principle preserve no supercharges\footnote{at some specific values of coupling some long multiplet could accidently shorten.}. 
Then, their scaling dimension is not protected by supersymmetry which therefore is a non-trivial function of the coupling $g$.

The CFT multiplets obey OPE selection rules. The one that will be relevant in our setup is the following \cite{Liendo:2018ukf}
\beq\label{eq:OPEfusion}
\mathcal{B}_1 \times \mathcal{B}_1  = \mathcal{I} + \mathcal{B}_2 + \sum_{\Delta>1} \mathcal{L}_{0,[0,0]}^{\Delta} \;,
\eeq
where $\mathcal{B}_1$ is the multiplet containing $\Phi_\perp^M$ as we discuss below. The $1/2$-BPS multiplets $\mathcal{B}_1$ and $\mathcal{B}_2$ were also considered in \cite{Giombi:2017cqn,Liendo:2016ymz,Cooke:2017qgm,Liendo:2018ukf,Ferrero:2021bsb}, $\mathcal{I}$ is the identity multiplet and ${\cal L}_{0,\[0,0\]}^\Delta$ are the non-protected long multiplets transforming as singlets under the global ${\rm SO}(5)\times{\rm SO}(3)$ symmetry.

Operators appearing in the sum of the r.h.s. of \eqref{eq:OPEfusion}, such as for example $\Phi_{||},\;\Phi_{||}^2$, $(\Phi_{\perp}\cdot \Phi_{\perp}),\;\dots$ mix and develop anomalous dimensions that can be computed using the QSC as shown in \cite{Grabner:2020nis,Cavaglia:2021bnz,Cavaglia:2022qpg,Julius:2021uka}. 
 By the unitarity bounds, it is known that these non-protected operators are all irrelevant, i.e. with $\Delta>1$~\cite{Agmon:2020pde}. 
Importantly, while $\Phi_{\perp}^M$ are protected exactly marginal operators with dimension $\Delta = 1$, the operator corresponding to $\Phi_{||}$ is an irrelevant operator with a running dimension $\Delta > 1$ (except exactly at zero coupling, where $\Delta \rightarrow 1$). 

Let us describe the structure of the $\mathcal{B}_1$ multiplet in more detail. 
It is the simplest protected multiplet and it contains three operators with the following quantum numbers
\beq\label{B1qn}
\mathcal{B}_1:\qquad\{1,\[0,1\],0\}\;\longrightarrow\;
\{\tfrac{3}{2},\[1,0\],\tfrac{1}{2}\}\;\longrightarrow\;
\{2,\[0,0\],1\}\,,
\eeq
where arrows represent the action of supercharges on the highest weight. This multiplet is also known as \textit{displacement multiplet}. Indeed, every defect theory has a distinguished operator called displacement operator, which captures the breaking of translation invariance by the defect. In particular, the stress-energy tensor is no longer conserved and the usual conservation law needs to be modified by some additional terms localised on the defect. In our case it leads to
\beq\label{Tmunu}
\partial_\mu T^{\mu n}=\mathbb{D}^n(x_{||})\delta^{3}(x_{\perp})\,,
\eeq
with $x_{||}=t$ the direction along the defect and $x_\perp$ the orthogonal ones. The operator $\mathbb{D}^n$ contains the components of the field-strength and it is protected. Given the definition \eqref{Tmunu}, its dimension is 2 and it corresponds to the last operator appearing in the multiplet  $\mathcal{B}_1$ \eqref{B1qn}. Also the R-current can be broken giving rise to other operators. These operators have protected dimension 1 and, in our case, they are just the marginal operators $\Phi_{\perp}^{M}$ with $M = 1,\dots, 5$ corresponding to the highest weight of the supermultiplet $\mathcal{B}_1$ \eqref{B1qn}.

\paragraph{Correlation functions.} Correlation functions are defined as local insertions of operators along the contour as follows \cite{Drukker:2006xg}
\begin{equation}\label{npoint}
    \left\langle\left\langle O_{1}\left(t_{1}\right) O_{2}\left(t_{2}\right) \cdots O_{n}\left(t_{n}\right)\right\rangle\right\rangle
    \equiv 
    \frac{\langle
    \operatorname{Tr}
   \operatorname{P}{ O}_1(t_1)
 \,\mathcal{W}_{t_1,t_2}\,{ O}_{2}(t_{2})
 \,   \ldots 
\, O_n(t_n)\, \mathcal{W}_{t_n,t_1}
    \rangle}{\langle  \mathcal{W}_{\mathcal{C}} \rangle}
    \; ,
\end{equation}
where $O_i$ are composite fields transforming in the adjoint representation of the gauge group and $\mathcal{W}_{t_a,t_b}$ are segment of the Wilson loop $\mathcal{W}_{\mathcal{C}}$ between positions $t_a$ and $t_b$. The double brackets $\left\langle\left\langle  \cdots \right\rangle\right\rangle$ indicate that the vev is taken with the supersymmetric Wilson loop as the vacuum instead of the usual one. 

The $n$-point functions \eqref{npoint} satisfy all the properties of a 1D conformal field theory. Then, two- and three-point functions are completely fixed by conformal symmetry. For instance, for scalar operators $O_{\Delta_i}$ with dimension $\Delta_i$ we have
\begin{equation}\begin{split}\label{2pt3ptCFT}
 \langle \langle O_{\Delta_i}(t_1) \;O_{\Delta_j}(t_2) \rangle\rangle &= N_i\frac{ \delta_{ij}}{x_{12}^{2 \Delta_i}}\equiv N_i\delta_{ij} [P(t_1,t_2)]^{\Delta_i},\\
  \langle \langle O_{\Delta_i}(t_1)\; O_{\Delta_j}(t_2)\;O_{\Delta_k}(t_3) \rangle\rangle &= \sqrt{N_iN_jN_k}\frac{ C_{ijk}}{x_{12}^{\Delta_i+\Delta_j-\Delta_k}\;x_{23}^{-\Delta_i+\Delta_j+\Delta_k}\;x_{13}^{\Delta_i-\Delta_j+\Delta_k}},
\end{split}\end{equation}
with $x_{ab}=|x^\mu(t_a)-x^\mu(t_b)|$ and $C_{ijk}$ the structure constant. 
We also introduced the function $P(t_1,t_2)=\frac{1}{|x(t_1)-x(t_2)|^2}$ for convenience.
We will use the two standard parametrisations: on the line parametrised by $x^\mu(t)=\{t,0,0,0\}$, $t\in [-\infty,+\infty]$ and on the unit circle parametrised by $x^\mu(t)=\{\cos t,\sin t,0,0\}$, $t\in [0,2\pi]$. So that we have $x_{ab}=|t_b-t_a|$ on the line and
$x_{ab}=\sqrt{2-2\cos (t_b-t_a)}$ for the circle.
We assume that by default in the first line of \eq{2pt3ptCFT} $N_i=1$, which is the standard normalisation of the operators. However, for the operators in the displacement multiplet \eqref{B1qn} it is more convenient to introduce a non-trivial normalisation as follows
\beq\label{2pt}
\langle \langle \Phi_{\perp}^M(t_1) \Phi_{\perp}^N(t_2)\rangle\rangle = \frac{ N_{\Phi_{\perp}} \,\delta^{MN}}{x_{12}^2},\qquad
\langle \langle \mathbb{D}^n(t_1) \;\mathbb{D}^m(t_2) \rangle\rangle = \frac{ N_{\mathbb{D}} \,\delta^{nm}}{x_{12}^4},\qquad
\eeq
where $C_{\Phi_{\perp}}$ and $C_{\mathbb{D}}$ are functions of the coupling as defined below in \eqref{normalisation}.

Conformal symmetry constrains also 4-point functions. Indeed, in one dimension they can be written in terms of a single cross ratio 
\beq\label{crossratio}
x=\frac{x_{12} x_{34}}{x_{13} x_{24}}\,,
\eeq
as follows
\beq\label{Gx}
{G}^{MNPQ}(x)\equiv\frac{\langle \langle \Phi_{\perp}^M(t_1) \Phi_{\perp}^N(t_2) \Phi_{\perp}^P(t_3) \Phi_{\perp}^Q(t_4)\rangle\rangle}{P(t_1,t_2)P(t_3,t_4)}\;,
\eeq
where $M,N,P,Q=1,\dots,5$. 
Exploiting the  superconformal OPE \eqref{eq:OPEfusion}, it is possible to parametrise the 4-point function \eqref{Gx} as
\beq\label{Gwithdelta}
G^{MNPQ}(x)=\delta^{MP}\;\delta^{NQ}\;G_2(x)+\delta^{MN}\;\delta^{PQ}\;G_1(x)+\delta^{MQ}\;\delta^{NP}\;G_3(x)\,,
\eeq
with
\beq\begin{split}\label{G1G2G3}
G_1(x)&=(x-1)f'(x)+\left(\frac{2}{x}-1\right)f(x)\,,\\
G_2(x)&=\mathbb{F} \;x^2-(x-1)xf'(x)-f(x)\,,\\
G_3(x)&=f(x)-xf'(x)\,,
\end{split}\eeq
and, in case of identical protected operators polarised in the same direction i.e. $M=N=P=Q$ as
\beq\begin{split}\la{pt4}
{G}(x) &=G_1(x)+G_2(x)+G_3(x)\\
&= \mathbb{F} \;x^2 +  (2 x^{-1} - 1)f(x) -\left(x^2 - x +1\right)f'(x)\; ,
\end{split}\eeq
where the reduced correlator $f(x)$ appearing in \eqref{G1G2G3} and \eqref{pt4} is a function of the cross ratio containing the OPE decomposition
\begin{equation}\label{eq:OPEf}
f(x) = F_{\mathcal{I}}(x) + { {C^2_{\rm BPS} \,  {F}_{\mathcal{B}_2}(x)}}  + \sum_{n } { {C^2_{n} \,  {F}_{{\Delta_n}}(x)}} \; ,
\end{equation}
with superconformal blocks given by 
\beqa\la{superblocks}
F_{\mathcal{I}}(x) &=& x\;,\\
F_{\mathcal{B}_2}(x) &=& x - x\, _2F_1(1,2,4;x ) \;,\\
F_{{\Delta}}(x) &=& \frac{x^{\Delta+1}}{1-\Delta}\, _2F_1(\Delta+1,\Delta+2,2 \Delta+4;x )\; ,
\eeqa
and OPE coefficients 
\beq\label{cn}
C_n \equiv C_{\Phi_{\perp}^i ,\; \Phi_{\perp}^i ,\;\mathcal{L}_{0,[0,0]}^{\Delta_n}}\,,
\eeq
for the non-protected states. Several orders of the reduced correlator are known in perturbation theory, at strong coupling in \cite{Ferrero:2021bsb} and at weak coupling in \cite{Kiryu:2018phb,Cavaglia:2022qpg}. 

The constant $\mathbb{F}$ appearing in \eqref{eq:OPEf} and the structure constant corresponding to the $\mathcal{B}_2$ block are related as $\mathbb{F}(g) = 1 + C^2_{BPS}(g)$. They can be computed exactly both using supersymmetric localisation \cite{Giombi:2017cqn,Liendo:2018ukf} or, alternatively, by making contact with  integrability with arguments similar to those of this paper, see  \cite{Cavaglia:2022qpg}.\footnote{See appendix F of the arXiv version.}
The result is
\beq\begin{split}\label{FF}
\mathbb{F}(g) = 1 + C^2_{BPS}(g) &=\frac{3\;\langle\mathcal{W}_{\text{circle}}\rangle\;\langle\mathcal{W}_{\text{circle}}\rangle''}{(\langle\mathcal{W}_{\text{circle}}\rangle')^2}\\
&= \frac{3 I_1(4 g \pi ) \left(\left(2 \pi ^2 g^2+1\right) I_1(4 g \pi )-2 g \pi  I_0(4 g \pi )\right)}{2 g^2 \pi ^2 I_2(4 g \pi ){}^2} \;\;, 
\end{split}\eeq
where the first expression refers to the expectation value of the circular Wilson loop given in \eqref{WLcircle}.

Finally, given the invariance under the cyclic relabelling of the 4-point function \eqref{Gx}, the quantities $G(x)$ and $f(x)$ satisfy the following crossing equations
\beq\begin{split}\label{crossingsym}
x^2 {G}(1 - x) - (1-x)^2 {G}(x) &= 0\,,\\
x^2 f(1 - x) + (1-x)^2 f(x) &= 0\; .
\end{split}\eeq
The main goal of this paper is to relate the deformation of this CFT with the integrability data. We discuss the deformations in the next section.

\subsection{The defect deformations}\label{sec:deformations}

\paragraph{The displacement deformations.} The Ward identity \eqref{Tmunu} fixes the variation of an arbitrary correlation function when the
contour of the defect undergoes a small deformation. Let us consider the deformation of a linear defect parametrised by $x_{||}=t$ by a profile $\delta x_\perp^n(t)$. Then, the correlation function $\langle\langle O\rangle\rangle$ of an arbitrary operator $O$ taken in presence of the deformed Wilson loop, at first order in the deformation reads \cite{Polyakov:2000ti,Semenoff:2004qr,Correa:2012at} 
\beq
\langle\langle \,O\,\rangle\rangle_{x+\delta x}=\int dt\;
\langle\langle \,O\;\mathbb{D}^n(t)\,\rangle\rangle\;
\delta x_{\perp}^n(t)+\mathcal{O}(\delta x^2)\,,
\eeq
where we assume that $\delta x(t)$ vanishes at the locations of the operators $O$. Similarly for the vev of the Wilson loop itself, with the difference that the first order variation vanishes since it corresponds to a one-point function on the defect CFT. The first non-trivial contribution appears at second order in the deformation and it is given by
\beq\label{dispdx}
\delta_x\log\langle \,\mathcal{W}_\mathcal{C}\,\rangle=\int_{t_1>t_2} dt_1dt_2\;
\langle\langle \,\mathbb{D}^n(t_1)\;\mathbb{D}^m(t_2)\,\rangle\rangle\;
\delta x_{\perp}^n(t_1)\delta x_{\perp}^m(t_2)+\mathcal{O}(\delta x_\perp^3)\;.
\eeq
For the supersymmetric Wilson loop \eqref{eqn:MWLdefine} there is also an internal angle displacement operator corresponding to $\Phi_{\perp}^N$, $N=1,\dots,5$, the highest weights of the displacement multiplet. These operators are sometimes referred to as \emph{tilt operators}. One can see these operators arising when  the coupling to the scalars in the Wilson loop connection  \eqref{eqn:MWLdefine} is deformed as $n+\delta n$, with $n\cdot\delta n=0$, leading to 
\beq\label{dispdn}
\delta_n\log\langle \,\mathcal{W}_\mathcal{C}\,\rangle=\int_{t_1>t_2} dt_1dt_2\;
\langle\langle \,\Phi_{\perp}^N(t_1)\;\Phi_{\perp}^M(t_2)\,\rangle\rangle\;
\delta n^N(t_1)\delta n^M(t_2)+\mathcal{O}(\delta n^3)\;.
\eeq

\paragraph{Integrability data for the WL deformations.}

\begin{figure}
    \centering
    \includegraphics[scale = 1]{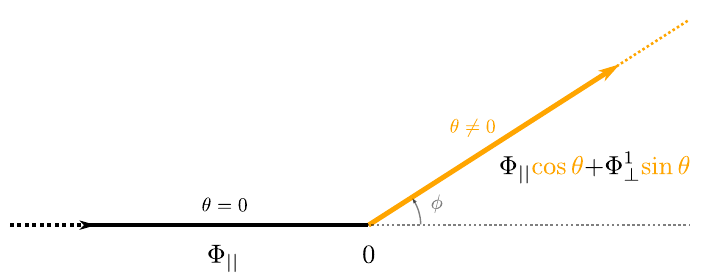}
    \caption{The WL deformed by introducing a point-like parameter changing defect is integrable. We utilise this fact to deduce further constraints on the correlation functions of the 1D CFT on the straight line.}
    \label{fig:cusp}
\end{figure}
A particular deformation of the defect CFT is given by introducing a
cusp along the contour as in figure \ref{fig:cusp}, which can be considered as a point-like parameter changing defect inside WL or also as a particular case of a colour-twist operator introduced in \cite{Cavaglia:2020hdb}. The resulting operator is composed of two semi-infinite lines connecting in the origin as follows
\begin{align}\label{Wcusp}
    {\cal W}_{\text{cusp}} =\frac{1}{N} \operatorname{Tr}\left[ W_{-\infty}^0(0,0)
    W_{0}^{+\infty}(\phi,\theta)\right]
    \;,
\end{align}
where the second infinite segment is rotated both in space-time with angle $\phi$ forming the cusp, as well as in the space of scalar couplings with the internal angle $\theta$. Choosing planes for the rotations, the two rays can be written as follows
\begin{equation}\label{eqn:thetaphinotation}
    {W}_{t_1}^{t_2}(\phi,\theta) = \operatorname{P}\exp\int_{t_1}^{t_2}dt\bigg[i\, A_\mu \dot{x}^{\mu}(t) + (\Phi_{||}\cos\theta + \Phi^1_\perp\sin\theta)\,|\dot{x}(t)|\bigg]\;,
    \end{equation}
with the contour parametrised by $x(t)= \big(t\cos\phi,t\sin\phi,0,0\big)$.

\begin{figure}
	\centering
	\includegraphics[width=\columnwidth]{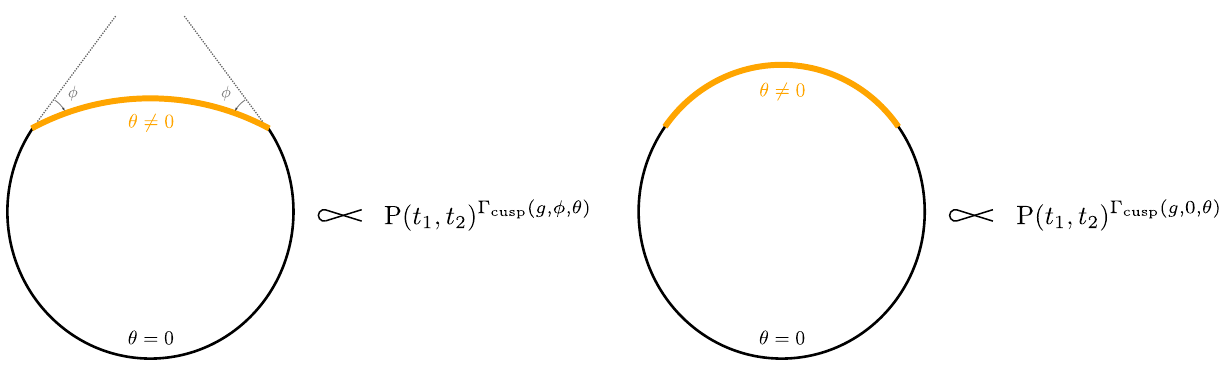}
	\caption{Another conformally equivalent representation of the cusped WL. The expectation value of this object depends on the distance between the cusps in the standard way as a two point function of two operators of dimension $\Gamma_{\rm cusp}$.}
	\label{fig:segment}
\end{figure}

The cusped Wilson line is no longer finite and it develops an anomalous dimension known as \emph{cusp anomalous dimension}.
This quantity is defined through the divergence of the vev of $\mathcal{W}_{\text{cusp}}$ as follows \cite{Korchemsky:1985xj}
\begin{align}\la{Wcusp2}
   \langle\,{\cal W}_{\text{cusp}}\,\rangle\sim \left(\epsilon_{\rm UV}\right)^{\Gamma_{\text{cusp}}(g,\phi,\theta)}\;,
\end{align}
where
$\epsilon_{\rm {UV}}$ is the UV cutoff near the cusp. Perhaps more transparent way of defining the cusp anomalous 
dimension, is by mapping the two lines to two intercepting arches of circles, meeting at the external angle $\phi$ as on figure~\ref{fig:segment}. The two arches has two interception points (one is the image of $0$ another is the image of $\infty$), then the expectation value will scale as the distance between these two intersection points (in 4D) to the power $-2\Gamma_{\text{cusp}}(g,\phi,\theta)$.

The cusp anomalous dimension $\Gamma_{\text{cusp}}(g,\phi,\theta)$ was introduced and studied  at weak and strong coupling in~\cite{Drukker:2011za}. Moreover, since 
the configuration introduced in \eqref{Wcusp} was discovered to be integrable, a set of TBA equations for it was introduced  in~\cite{Correa:2012hh,Drukker:2012de} and reformulated in terms of the QSC in~\cite{Gromov:2015dfa} allowing for its non-perturbative analysis.

In the near-BPS limit $\phi\to\pm\theta$, the first few orders of the cusp anomalous dimension are given by 
\beq
\label{dexpphi}
	\Gamma_{\text{cusp}}(g,\phi,\theta)=\frac{\cos\phi-\cos\theta}{\sin\phi}
	\frac{2\phi}{1-\frac{\phi^2}{\pi^2}} \mathbb{B}(g,\phi)
	+\left(\frac{\cos\phi-\cos\theta}{\sin\phi}\right)^2\phi^2 \mathbb{C}(g,\phi)+...\;,
\eeq
where the dots represents higher orders in $\tfrac{\cos\phi-\cos\theta}{\sin\phi}$. 
The function $\mathbb{B}(g,\phi)$ is known as \textit{Bremsstrahlung function} originally computed in \cite{Correa:2012at,Fiol:2012sg} exploiting the supersymmetric localisation techniques~\cite{Erickson:2000af,Drukker:2000rr,Drukker:2006ga,Pestun:2009nn}.  The same result was later reproduced and further generalised from integrability in  \cite{Gromov:2012eu,Gromov:2013qga} and checked at strong and weak coupling in \cite{Sizov:2013joa,Bonini:2015fng}. The function $\mathbb{C}(g,\phi)$ appearing in \eqref{dexpphi} was computed analytically in \cite{Gromov:2015dfa} using the QSC formalism. We will refer to it as \emph{Curvature function}.

For what follows it is useful to consider the cusp anomalous dimension in the case in which the Euclidean angle $\phi$ is set to zero and $\theta$ is small. This corresponds to study a small deformation of the $1/2$-BPS Wilson line,  leading to the following expansion
\begin{equation}\label{gammasin}
    \Gamma_{\text{cusp}}(g,\phi=0,\theta\rightarrow 0)=
     \mathbb{B}(g)\,\sin^2\theta
    +\frac{1}{4}\left(\mathbb{B}(g)+\mathbb{C}(g)\right)\,\sin^4\theta +\mathcal{O}(\sin^6\theta)\;.
\end{equation}
Finally, let us write the expression for the Bremsstrahlung and  Curvature functions
\begin{align}
   \label{bremdef}\mathbb{B}(g) &= \frac{g}{\pi}\frac{I_2(4 \pi g)}{I_1(4 \pi g)}\;,\\
    \label{curvaturedef}\mathbb{C}(g) &= -4\,\mathbb{B}^2(g)
	-\frac{1}{2}\oint\frac{du_x}{2\pi i}\oint\frac{d
 u_y}{2\pi i}
    K_0(u_x-u_y)F[x, y]\;,
\end{align}
where we use the shorthand notation $\mathbb{B}(g,\phi=0)\equiv \mathbb{B}(g)$ and $\mathbb{C}(g,\phi=0)\equiv \mathbb{C}(g)$.
Here both integrals run clockwise around the cut $[-2g,2g]$ and $u_x = g ( x + 1/x)$ is the Zhukovsky parametrisation. The kernel $K_0$ and the integrand $F$ are given in appendix \ref{app:curvature}. 

Curiously, the function $\mathbb{F}$, appeared in the correlator \eqref{FF} could be written in terms of the $\mathbb{B}$ as follows
\beq\label{eq:F0}
\mathbb{F}(g) =  \frac{3 (g^2 -\mathbb{B}(g))}{\pi^2 \mathbb{B}(g)^2}\;\;.
\eeq
The integral \eqref{curvaturedef} was solved perturbatively in \cite{Cavaglia:2022qpg}. At weak coupling it gives
\beq\begin{split}\label{eq:weakC0}
\mathbb{C}&=
4 g^4-\left(24 \zeta_3+\frac{16 \pi ^2}{3}\right)g^6+ \left(\frac{64 \pi ^2 \zeta_3}{3}+360 \zeta_5+\frac{64 \pi ^4}{9}\right)g^8\\
&- \left(\frac{112 \pi ^4 \zeta_3}{5}+272 \pi ^2 \zeta_5+4816 \zeta_7+\frac{416 \pi ^6}{45}\right)g^{10}\\
&+\left(\frac{3488 \pi ^6 \zeta_3}{135}+\frac{2192 \pi ^4 \zeta_5}{9}+\frac{9184 \pi ^2 \zeta_7}{3}+63504 \zeta_9+\frac{176 \pi
  ^8}{15}\right)g^{12} +\mathcal{O}\left(g^{14}\right)\;.
\end{split}
\eeq
At strong coupling, the coefficients of the series in $g$ were deduced by evaluating the integral with very high numerical precision and then fitting with Riemann zeta values $\zeta_n$ obtaining the following expansion
\beq\begin{split}
\mathbb{C}&=
\frac{\left(2 \pi ^2-3\right) g}{6 \pi ^3}+\frac{-24 \zeta_3+5-4 \pi
   ^2}{32 \pi ^4}+\frac{11+2 \pi ^2}{256 \pi ^5 g}+\frac{96 \zeta_3+75+8 \pi ^2}{4096 \pi ^6 g^2}\\
   &+\frac{3 \left(408 \zeta_3-240
   \zeta_5+213+14 \pi ^2\right)}{65536 \pi ^7 g^3}+\frac{3 \left(315
   \zeta_3-240 \zeta_5+149+6 \pi ^2\right)}{65536 \pi ^8
   g^4}+\mathcal{O}\left(\frac{1}{g^5}\right)\;.\label{eq:strongC}
\end{split}
\eeq
for the next $3$ coefficients we were only able to find their numerical values
\beq
\frac{3.044012903724157826\times 10^{-7}}{g^5}+
\frac{8.008516278599531\times 10^{-8}}{g^6}+
\frac{2.125834835083\times 10^{-8}}{g^7}\nonumber
\eeq
In the ancillary file of this paper, we included the {\it Mathematica} notebook \texttt{Curvature.nb} containing the implementation we used to numerically compute 
the curvature function with high precision. 

\paragraph{From cusp anomalous dimension to  two-point function.}
Let us relate the following two quantities: the normalisation of the two-point function of the deformation operators, and the cusp anomalous dimension at the leading non-trivial order in the deformation parameter following~\cite{Correa:2012at}. The rest of the paper will be dedicated to generalising this consideration to the higher order in the deformation parameter, so this example, while simple, is very useful to demonstrate the general idea.
Consider, first, the deformation by $\theta$ while keeping $\phi=0$.
We introduce non-zero $\theta$ in the interval $[-T/2,T/2]$, which can be interpreted as a two-point function of two defects and thus we should have
\beq\la{Wtheta}
\mathcal{W}_\mathcal{C}\simeq
(T/\epsilon)^{-2\Gamma_{\rm cusp}}
\simeq 1-2\theta^2{\mathbb B}\log\frac{T}{\epsilon}
,
\eeq
where $\epsilon$ is a UV cut-off 
and we used that
 from \eq{dexpphi} $\Gamma_{\rm cusp}=\theta^2 {\mathbb B}$
.
Our starting point is \eq{dispdn} with 
$\delta n(t)=(0,0,0,0,\theta,0)$ for $t\in [-T/2,T/2]$ and zero otherwise
\beq\label{dispdx2}
\delta_x\log\langle \,\mathcal{W}_\mathcal{C}\,\rangle=
\int_{-T/2}^{T/2-\epsilon} dt_1
\int_{t_1+\epsilon}^{T/2} dt_2
\;
\frac{ N_{\Phi_{\perp}}}{(t_1-t_2)^2}
\;
\theta^2+{\cal O}(\theta^3)\;.
\eeq
Above we introduced the UV cut-offs $\epsilon$. The integration can be evaluated exactly
\beq
\delta_x\log\langle \,\mathcal{W}_\mathcal{C}\,\rangle =\(
\frac{T}{\epsilon }-\log \frac{T}{\epsilon }-1\)C_{\Phi_{\perp}}\theta^2\,.
\eeq
In the section~\ref{sec:derive} we will introduce a more precise treatment of the linear divergence, but for now we can just drop it and comparing the $\log$ terms with 
\eq{Wtheta} to obtain 
$
N_{\Phi_\perp}=2{\mathbb B}
$.

Similarly, we can analyse the deformation by $\phi$. The small deformation by $\phi$ is equivalent to introducing a tiny bump in the line between $(-T/2,T/2)$, which can be written at the linear order as a parabolic deformation $\delta x(t)=\frac{\phi}{T}\(\frac{T^2}{4}-t^2\)$, approximating a tiny arc of a circle inclusion. One can check that indeed the slope of this line at $t=\pm T$ is $\pm\phi$. From \eq{dispdx} we get
\beq\label{dispdx3}
\delta_x\log\langle \,\mathcal{W}_\mathcal{C}\,\rangle=
\int_{-T/2}^{T/2-\epsilon} dt_1
\int_{t_1+\epsilon}^{T/2} dt_2
\;
\frac{ N_{\mathbb{D}} }{(t_1-t_2)^4}
\delta x(t_1)\delta x(t_2)
+{\cal O}(\phi^3)\;,
\eeq
the log-divergent part of the integral is $\frac{1}{6}C_{\mathbb{D}}\phi^2\log \frac{T}{\epsilon}$
again comparing with $\Gamma_{\rm cusp}=-\phi^2 {\mathbb B}$, as follows from \eq{dexpphi} we get $N_{\mathbb{D}}=12 \mathbb{B}(g)$. To summarise we get the following normalisations of the 2-point functions
\beq\label{normalisation}
N_{\mathbb{D}}=12 \mathbb{B}(g)\qquad\text{and}\qquad
N_{\Phi_\perp}=2\mathbb{B}(g)\;.
\eeq
Next, we remind the form of the integrated correlators initially found in \cite{Cavaglia:2022qpg} with the goal to derive a combination of them in the rest of the paper.

\subsection{Integrated correlators}\label{sec:integrated0}

In the previous section we discussed how the deformation of the line by a cusp at the leading order produces a normalisation of the two point function. Already at the next order we get a nontrivial relation which involves an integrated four-point function. We will consider the simplest four-point function of four identical protected operators $\Phi_\perp^N$ introduced in \eqref{Gx}, which will be related to the R-symmetry deformation of the Wilson loop operator at next-to-leading order \eqref{gammasin}.    
At this order, one can write the following independent constraints \cite{Cavaglia:2022qpg}, expressed in terms of the reduced correlator $f(x)$ defined in \eqref{eq:OPEf}, 
\begin{align}
  &\text{ Constraint 1:  }&&   \int_{\delta_x}^1
     \delta f(x)\( 
    \frac{1}{x}+\frac{1}{x^3}
    \)dx
    -\frac{1}{2} (\mathbb{F}-3) \log {\delta_x}-\mathbb{F}+3
    =\frac{3\mathbb{C}-\mathbb{B}}{8 \;\mathbb{B}^2}\; , \label{eq:constr1}\\
   & \text{ Constraint 2:  }&& \int_{0}^1  \frac{\delta f(x)}{x}dx = 
        \frac{\mathbb{C}}{4\;\mathbb{B}^2} + \mathbb{F}-3 \; ,\label{eq:constr2}
\end{align}
where $\delta_x\rightarrow 0^+$ is a cut-off regulator and $\delta f(x) \equiv f(x) - f_{\text{tree}}(x)$ with
$f_{\text{tree}}$ the zero-coupling value
\beq\label{ftree}
f_{\text{tree}}(x)=2 x+\frac{x}{x-1}\; .
\eeq
Notice that the expression in the l.h.s. of \eqref{eq:constr1} is finite since $\delta f(x)\sim\frac{3-\mathbb{F}}{2}x^2$ for $x\rightarrow 0$. The Bremsstrahlung ${\mathbb B}$ and Curvature functions ${\mathbb C}$ are defined in \eqref{bremdef} and \eqref{curvaturedef}. 

For completeness, it is worth mentioning that  one of the integrals appearing in \eqref{eq:constr1} and \eqref{eq:constr2} can be further simplified by
\beq
\int_0^1\frac{\delta f(x)}{x}\,dx=\int_0^1\delta f(x) \,dx\,,
\eeq
as follows immediately from the crossing relation \eq{crossingsym}, which implies that $f(x)(1-1/x)$ is an odd function under crossing and thus vanishes under the integral.

Exploiting the relation between the 4-point function $G(x)$ and the reduced correlator $f(x)$ given in \eqref{pt4}, the first integrated constraint \eqref{eq:constr1} can also be written as follows
\beq
\text{ Constraint 1:  }  \int_0^1
    \delta G(x)\frac{1+\log x}{x^2}dx=\frac{3\mathbb{C}-\mathbb{B}}{8 \;\mathbb{B}^2}\; , \label{eq:constr1old}
\eeq
where $
\delta G(x) \equiv G(x) - G_{\text{tree}}(x)$ and $G_{\text{tree}}$ is the zero-coupling value:
\beq\label{Gtree}
G_{\text{tree}}(x) = \frac{2 (x-1) x+1}{(x-1)^2}
\; .
\eeq

The integral relations \eqref{eq:constr1} and \eqref{eq:constr2} were tested for several orders at weak and strong coupling in \cite{Cavaglia:2022qpg}. Recently, the authors of \cite{Drukker:2022pxk} managed to derive a linear combination of these constraints by using a general argument on the geometry of the conformal manifold. The combination derived in \cite{Drukker:2022pxk} can be written as 
\beq\label{Nadav}
-2\int_{\delta_x}^1
     \delta f(x)\( 
    \frac{1}{x}+\frac{1}{x^3}
    \)dx+(\mathbb{F}-3) \log {\delta_x}+3\int_{0}^1  \frac{\delta f(x)}{x}dx=\frac{1}{4\mathbb{B}}+\mathbb{F}-3\,,
\eeq
and is independent from the Curvature function. In \cite{Drukker:2022pxk} it was derived by studying the Riemann tensor of the defect conformal manifold generated by the marginal operator $\Phi_\perp$. The aim of this paper is to provide the proof of the other independent linear combination of the constraints.

\section{Strategy and Technicalities}\label{sec:CPT}

In this section we describe the main strategy of our derivation, and then proceed to describe in detail the conformal perturbation theory setup we use for the calculation. The most technical parts of the derivation will be described in the next section.

\subsection{The main strategy}

Our main goal is relating the cusp anomalous dimension to the integrated correlator of the four scalar operators $\Phi_\perp$. The latter, as has been already discussed, are the ``tilt operators'' controlling deformations of the internal angle along the MWL. Therefore, we will consider the ``cusp'' defined purely in  R-space, i.e. we only consider the deformation by the angle $\theta$ in the notation of Section \ref{sec:deformations}. In this case,  $\Gamma_{\text{cusp}}$ can be viewed as a scaling dimension of two point-like parameter-changing defects, see the right panel of  Figure \ref{fig:segment}. Namely, we consider a straight (or circular) MWL, where the polarisation of the scalars is modified in an interval between points $t_1$ and $t_2$. This configuration has an expectation value of the form:
\beq\label{eq:2cusptheta}
 {\cal W}(t_1,t_2;\theta) \equiv\frac{ \langle W_{-\infty}^{t_1}(0,0)\, W_{t_1}^{t_2}(0, \theta) \, W_{t_2}^{\infty}(0,0) \rangle }{\langle W_{-\infty}^{\infty}(0,0)\rangle} \propto \left(\epsilon_{\text{UV}}^2/x_{12}^2 \right)^{\Gamma_{\text{cusp}}(0,\theta) } ,
\eeq
where $\epsilon_{\text{UV}}$ is a UV cutoff controlling the neighbourhood of the points $t_1$ and $t_2$, as we discussed around \eq{Wcusp2}. This formula follows from the definition of $\Gamma_{\text{cusp}}$ and conformal invariance in the SYM theory, see e.g. \cite{Cavaglia:2018lxi,Dorn:2020meb}.

The above expression is written in terms of  quantities in $\mathcal{N}$=4 SYM. Our first goal is to rewrite the small $\theta$ expansion of this observable in terms of internal quantities of the 1D defect CFT living on the straight Wilson line. We first explain the main intuition in a treatment suitable at weak coupling, and later we will explain the proper CFT setup suitable to keep the 't Hooft coupling finite.

\subsubsection{From deformations of the MWL to integrated correlators}

It is a general expectation that deformations of a Wilson line can be realised in terms of a series of integrated operator insertions on the undeformed contour  \cite{Drukker:2006xg,Cooke:2017qgm}. 

In our case, this can be seen easily from the definition (\ref{eqn:thetaphinotation}), which we repeat here for convenience in the case $\phi = 0$ and assuming a straight MWL along the direction $4$ in spacetime
\beq\label{eq:action}
 {W}_{t_1}^{t_2}(0,\theta) = \operatorname{P}\exp\int_{t_1}^{t_2}dt\bigg[i\, A_4(t)  + (\Phi_{||}(t) \cos\theta + \Phi^1_\perp(t) \sin\theta)\bigg]\; .
\eeq
Considering $\mathbf{s} \equiv \sin\theta$ a small expansion parameter, expanding \eq{eq:action} up to the ${\bf s}^4$ we find
\beqa\label{eq:action2}
 {W}_{t_1}^{t_2}(0,\theta) &\sim& \operatorname{P} e^{\int_{t_1}^{t_2}dt\left( i\, A_3(t)  + \Phi_{||}(t) \right) }\times\exp\int_{t_1}^{t_2}dt \;\delta L\;\;,\\
\nn \text{with}\quad\delta L&=&\bigg[ \mathbf{s}\; \Phi_{\perp}^1(t) -\frac{\mathbf{s}^2}{2}\; \Phi_{||}(t) - \frac{\mathbf{s}^4}{8} \;\Phi_{||}(t) + \dots \bigg]\,.
\eeqa
We found that in order to get a finite result we have to add divergent counter-terms given by insertions of the identity operator.\footnote{Note that, the identity is the the only \emph{relevant} operator of the 1D CFT, i.e. with $\Delta < 1$~\cite{Agmon:2020pde}.} So, in practice we should use
\beqa\la{dLren}
\nn \delta L&=&\bigg[ \mathbf{s}\; \Phi_{\perp}^1(t) +\mathbf{s}^2\(\frac{b_{2}}{\epsilon}-\frac{1}{2}\; \Phi_{||}(t)\)+\mathbf{s}^4\(\frac{b_{4}}{\epsilon}- \frac{1}{8} \;\Phi_{||}(t)\) +{\cal O}({\bf s}^6) \bigg]\,,
\eeqa
where $\epsilon$ is the UV regularisation, which will be described in detail in the next section.

For the left hand side of (\ref{eq:2cusptheta}) we get
\beqa\label{eq:expansion}
&& {\cal W}(t_1,t_2;\theta) \sim 1 
+ \mathbf{s}^2 {\Big [}\int\limits_{t_1<s_1<s_2<t_2} ds_1 \,ds_2 \, \langle \langle \Phi_{\perp}^1(s_1) \Phi_{\perp}^1(s_2) \rangle \rangle +\frac{b_2}{\epsilon} \int_{t_1}^{t_2} ds_1 \langle \langle \mathds{1}(s_1) \rangle \rangle  {\Big ] } \nonumber \\
&&+\frac{\mathbf{s}^4 }{4}  \, {\Big [ }\!\!\!\!\int\limits_{t_1<s_1<s_2<t_2} \!\!\!\!\!\!\!\!\!\!\!ds_1 \,ds_2 \, \langle \langle \Phi_{||}(s_1) \Phi_{||}(s_2) \rangle \rangle+\frac{4b_4}{\epsilon}\!\! \!\!\int\limits_{t_1<s<t_2} \!\!\!\!\!\!\!ds {\langle \langle \mathds{1}(s) \rangle \rangle}
+\frac{4b_2^2}{\epsilon^2}\!\!\!\! \int\limits_{t_1<s_1<s_2<t_2}\!\!\!\!\!\!\!\!\!\!\!ds_1ds_2
\, \langle \langle \mathds{1}(s_1)\mathds{1}(s_2) \rangle \rangle {\Big ] } \nn\\
&&-\frac{\mathbf{s}^4}{2} \int\limits_{t_1<s_1<s_2<s_3<t_2} ds_1 \,ds_2 \,ds_3 \left( \langle \langle \Phi_{||}(s_1) \Phi_{\perp}^1(s_2) \Phi_{\perp}^1(s_3) \rangle\rangle + \texttt{cyclic} \right) \\
&&+\frac{b_2\mathbf{s}^4}{\epsilon} \int\limits_{t_1<s_1<s_2<s_3<t_2} ds_1 \,ds_2 \,ds_3 \left( \langle \langle \mathds{1}(s_1) \Phi_{\perp}^1(s_2) \Phi_{\perp}^1(s_3) \rangle\rangle + \texttt{cyclic} \right) \nn\\
&&+\mathbf{s}^4\, \int\limits_{t_1<s_1<s_2<s_3<s_4<t_2} ds_1 \,ds_2 \,ds_3\,ds_4 \, \langle \langle \Phi_{\perp}^1(s_1) \Phi_{\perp}^1(s_2) \Phi_{\perp}^1(s_3) \Phi_{\perp}^1(s_4) \rangle\rangle \nn + {\cal O}\left( \mathbf{s}^6 \right)\,,
\eeqa
where we have already set to zero all correlators appearing in the expansion which vanish for R-symmetry reasons.

As reviewed in section \ref{sec:integrated0}, the order $\mathcal{O}(\mathbf{s}^2)$ of this equation, compared with the expansion of $\Gamma_{\text{cusp}}$, fixes the normalisation of the 2-point function of $\Phi_{\perp}^N$ in terms of the  Bremsstrahlung function $\mathbb{B}$, defined in \eqref{gammasin}. At the $\mathcal{O}(\mathbf{s}^4)$ order the last term gives an integrated 4-point function, which by comparing with the r.h.s. of  \eq{eq:2cusptheta} should be related to the curvature function ${\mathbb C}$.

Going to higher orders requires to be consistent and careful about regularisation, see e.g. \cite{Cooke:2017qgm} for studies at weak coupling. 
In order to deal with the regularisation scheme consistently at finite coupling, in the next section we describe a more abstract formalism of conformal perturbation theory. 

\subsubsection{Conformal perturbation theory framework}

In this section we describe a more abstract point of view about the deformation by the angle $\theta$, using exclusively the 1D CFT language rather than referring to fields in $\mathcal{N}$=4 SYM. 

Namely, we view the ${\bf s}=\sin\theta$-deformation as a deformation (locally) by a marginal operator $\Phi_{\perp}^1$ at the leading order, which is also accompanied with a perturbation by relevant (such as the identity   $\mathds{1}$) and irrelevant (such as $\Phi_{||}$ and possibly others) operators at higher orders. We then constrain the way the irrelevant operators appear by requiring that the deformation is a symmetry or rotation in R-space when extended to the whole line. As at the higher orders in the deformation parameter the regularisation  has to be applied consistently, we also give in this section the detailed description of our regularisation scheme.
Our consideration follows closely the general conformal perturbation theory framework
(see e.g. \cite{Cardy:1996xt,Zamolodchikov:1987ti,ZAMOLODCHIKOV1989641,Kutasov:1988xb,Ranganathan:1993vj,Gaberdiel:2008fn,Amoretti:2017aze,Behan:2017mwi,Baggio:2017aww,Gabai:2022vri} for examples of the method and its many applications).

\paragraph{The deformation. }
The abstract action of the 1D CFT at $\theta = 0$ will be denoted as  $\mathcal{A}_{CFT} $. Then we write the $\theta$-deformed action as a local action in terms of the operators of the undeformed CFT. 

Since $\theta$ does not break conformal invariance locally, the  perturbation at leading order in $\mathbf{s}$ should be driven by an exactly marginal operator (i.e. with $\Delta = 1$). The only such operators in the 1D CFT are the ``tilt operators'', and, compatible with the symmetries of the problem, we should have:
\beq
\mathcal{A}_{\text{CFT}}(\theta) \sim \mathcal{A}_{\text{CFT}}(0) + \mathbf{s}\; \int dt \; O_{\Phi^1_{\perp}}(t) + \mathcal{O}(\mathbf{s}^2)\;,
\eeq
where we use notations such as $O_{\Phi^1_{\perp}}(t)$ to denote the primary operator of in the 1D CFT corresponding to the insertion of the field $\Phi^1_{\perp}$ at weak coupling. We use this notation to make clearer the distinction between internal 1D CFT quantities and the ${\cal N}=4$ SYM fields. 
 At this order, clearly this is the same as (\ref{eq:action2}).
 
 The normalisation of the operator can be fixed to be related to the Bremsstrahlung function:
\beq
\langle O_{\Phi^1_{\perp}}(t_1)  O_{\Phi^1_{\perp}}(t_2) \rangle_{\oneD} = 2 \mathbb{B}\, {\rm P }(t_1,t_2)\; ,
\eeq
namely, this will ensure that the deformation parameter $\textbf{s}$ is correctly identified with $\sin\theta$, up to $\mathcal{O}(\theta^2)$ order. This could be done exactly as described before in \ref{sec:integrated0}, and here we incorporate it from the beginning to simplify the following steps in the derivation.

At higher orders in $\mathbf{s}$, the deformed action may also include relevant and irrelevant operators, as in general happens for exactly marginal deformations considered in conformal perturbation theory, see e.g. \cite{Gaberdiel:2008fn}. 
Fixing the couplings of all these coefficients, which can be expected to be heavily regularisation scheme-dependent, is in general complicated, but we will find a way to constrain them to the order that we are interested in. 

We  start by being completely agnostic and assume that the action at generic order in $\mathbf{s}$ takes the form
\beq\label{eq:actiondef}
\mathcal{A}_{\text{CFT}}(\theta) \sim \mathcal{A}_{\text{CFT}}(0) + \delta \mathcal{A}_{\text{CFT}}\,,
\eeq
with
\beq\label{eq:deltaA}
 \delta \mathcal{A}_{\text{CFT}} =\mathbf{s}\; \int dt \; O_{\Phi^1_{\perp}}(t) + \sum_{k=2}^{\infty} \; \mathbf{s}^{k} \;\sum_{n} b_{n,k}\; \epsilon^{\Delta_n - 1} \int dt \; O_{n}(t) .
\eeq
Above, $O_n(t)$ are all local operators in the initial CFT and $\Delta_n$ their scaling dimensions. The prefactor $\epsilon^{\Delta_n-1}$, where $\epsilon$ is a dimensional parameter, has to be introduced for dimensional reasons. As described in the next section, in our regularisation scheme $\epsilon$ will be identified with the UV cutoff.

For convenience, let us also define the ``Lagrangian density''
\beq~\label{eqn:deltaLdef}
\delta L(t) =\mathbf{s}\; O_{\Phi^1_{\perp}}(t) + \sum_{k=2}^{\infty} \; \mathbf{s}^{k} \;\sum_{n} b_{n,k}\; \epsilon^{\Delta_n - 1} \; O_{n}(t).
\eeq
We assume that the perturbation $\bf s$ can in general be a function of $t$ or at least piecewise constant, such as for the case of the cusp.  Comparing with \eq{eq:action2}, we can make the following assumptions: 1) $O_{\Phi^1_{\perp}}(t)$ only appears at the linear order in ${\bf s}=\sin\theta$, which defines the coupling and prevent it from possible redefinitions of the type ${\bf s}\to {\bf s}+\alpha {\bf s}^3+\dots$. Therefore, the sum on the r.h.s. of the equation will \emph{not} include $O_{\Phi^1_{\perp}}(t)$ or any of the other marginal operators $O_{\Phi^i_{\perp}}(t)$, but only the identity and irrelevant operators; 2) from ${\cal N}=4$ SYM considerations, it is natural to expect that in the sum \eq{eqn:deltaLdef} only the operator $O_{\Phi_{||}}(t)$ appears. However, we will lift this restriction, which may appear to be scheme dependent. As we will see, the possible existence of other operators in the sum does not affect our derivation, which requires only the assumption of point 1) above. 

Notice that, as discussed above, the identity operator will appear into the sum \eq{eqn:deltaLdef} as a counterterm, allowing  
us to remove consistently all power-like divergences, like those in the example at the end of section~\ref{sec:deformations}.

Finally, let us fix the following convention for the normalisation of the operators:  we incorporate from the beginning the normalisation for the tilt operators:
\beq
\langle O_{\Phi^i_{\perp}}(t_1) O_{\Phi^j_{\perp}}(t_2) \rangle_{\oneD} = \frac{2 \mathbb{B} \delta_{ij}}{t_{12}^2}\; ,
\eeq
while all other operators are normalised canonically:
\beq
\langle O_{n}(t_1) O_{m}(t_2) \rangle_{\oneD} = \frac{ \delta_{m n}}{t_{12}^{2 \Delta_n}}\;.
\eeq
\paragraph{Observables. }
The coefficients $b_n(\mathbf{s}) = \sum_{k=2}^{\infty} b_{n,k} \mathbf{s}^k$ are called Wilson coefficients and we will discuss later how to constrain them. 
Assuming we knew all of them,  observables in the $\theta$-deformed CFT are defined by the formal expansion
\beqa\label{eq:integralsL}
\langle \dots \rangle_{\theta} &=& \langle {\rm P}\dots e^{\int \delta L(t) dt }\rangle_{\oneD}
\\
\nn &=& 
\int dt\; 
\langle {\rm P} \dots \delta L(t) \rangle_{\oneD} + \int_{s_1<s_2} \langle {\rm P}\dots \delta L(s_1) \delta L(s_2) \rangle_{\oneD} + \dots,
\eeqa
where $\dots$ indicates possible operator insertions. This gives an expansion in terms of integrated correlators in the original CFT. The time/path ordering in this context is the standard prescription of the perturbation theory. 

\paragraph{Connection with the cusp. }
As explained above, we will obtain a nontrivial constraint on the CFT data by considering the deformation switched on only on the segment between two-points $t_1$ and $t_2$:
\beq\label{eq:constr0c}
 {\cal W}(t_1,t_2;\theta) \sim \langle {\rm P}e^{\int_{t_1}^{t_2} \delta L(t) dt }\rangle_{\oneD}\propto \left(\epsilon_{\text{UV}}^2/x_{12}^2 \right)^{\Gamma_{\text{cusp}}(0,\theta)}\; ,
\eeq
which connects us to the cusp anomalous dimension as in \ref{eq:2cusptheta}.

In the following section we explain in more detail the rules of conformal perturbation theory giving the details of our regularisation scheme. Then we proceed to discuss how we constrain the action, and introduce the concrete calculations which will be then presented in the next section.

\subsection{Regularisation scheme}\label{sec:regularisation}

\paragraph{Regularisation scheme. }
The integrals arising from the expansion (\ref{eq:integralsL}) will present short-distance as well as IR divergences. The IR divergences are easy to deal with: we will simply consider the theory on the circle of the circumference $2\pi$, i.e. in the 1D CFT internal terms we introduce finite temperature with the corresponding modification of the correlators with $t_{12}^{-2} \rightarrow {\rm P}(t_{1},t_2)=\frac{1}{2-2\cos(t_1-t_2)}$.
To regularise the UV divergences we introduce a hard point-splitting cutoff $\epsilon$ as follows
\begin{itemize}
    \item In all iterated integrals, we restrict integration variables $s_i$, $s_j$ so that $|s_i-s_j|>\epsilon$. 
    To be more explicit, we will enforce this by introducing integration measures 
    \beq\label{mun}
    \mu_n(s_1, \dots, s_n)
    \equiv \prod_{i<j}^n \Theta(|s_i-s_j|-\epsilon) ,
    \eeq
    (where $\Theta$ is the Heaviside step function),  so that the $n$-fold iterated integral is computed as
\beq
\int_{s_1<s_2<\dots <s_n} ds_1 \dots ds_n \; \mu_n
\; \langle \delta L(s_1) \dots \delta L(s_n) \rangle_\oneD\;.
\eeq
    \item When we consider operator insertions in non-integrated points $t_i$, we also restrict all integration variables to a distance $\epsilon$ from all operators.
    This can be recast into the following integration measure
    \beq\label{munm}
    \mu_{n,m}
    (s_1, \dots, s_n;t_1,\dots,t_m)
    \equiv \mu_n\;
    (s_1, \dots, s_n)
    \prod_{i=1}^n \prod_{j=1}^m \Theta(|s_i - t_j|-\epsilon) ,
    \eeq
    Again, explicitly, the integrals with insertions are computed as
    \beqa
    \int_{s_1<s_2<\dots <s_n} ds_1 \dots ds_n && \mu_{n,m}
    \;\langle O(t_1) \dots \delta  L(s_1) \dots \delta L(s_n) \dots O_m(t_m) \rangle_\oneD.
    \eeqa
    In the rest of the paper, in order to lighten the notation, we drop the explicit dependence on the points of the measures \eqref{mun} and \eqref{munm}.  
     \item $\epsilon$ is identified with the dimensional parameter appearing in the action \eq{eq:deltaA}. This fixes the convention for the Wilson's coefficients $b_{n,k}$. 
      \item The regularised value of observables is defined keeping all terms up to $\mathcal{O}(1)$ for $\epsilon \rightarrow 0$. In particular, we will tune the couplings so that divergences cancel and we are left with a finite result. 
      Notice that we should first keep $\epsilon$ finite (but much smaller than $2\pi$) both in the action and in the cutoffs, and that we send $\epsilon \rightarrow 0$ only on the final result for the correlator. 
\end{itemize}

\noindent
As a result of the above rules, we will see that we cannot throw away the terms corresponding to irrelevant operators in the action, which naively are suppressed in the action (\ref{eq:deltaA}) because they have $\Delta >1$. In fact, it will happen that, after integration inside correlation functions, these terms produce divergences that balance with the prefactor $\epsilon^{\Delta -1}$, giving a finite contribution.

\subsection{Constraints on the Wilson coefficients}
\begin{figure}
    \centering
    \includegraphics[width=0.7\columnwidth]{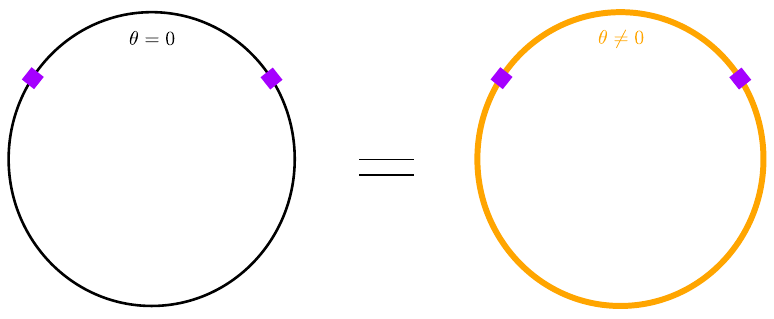}
    \caption{The figure illustrates one of the main conditions imposed in our derivation, where the thick orange line represents the line with the deformation switched on ($\theta \neq 0$), and the thin dark line represents the original 1D CFT with $\theta=0$. The inserted operators are $\frac{1}{2}$-BPS tilt operators with polarisation orthogonal to the plane of the $\theta$-rotation.   We demand that  such 2-point function should be invariant.}
    \label{fig:orthoInsert}
\end{figure}

Rather than fixing completely the higher-order coefficients, we will impose some physical conditions that will result into sum rules. The conditions we will exploit descend from the fact that $\theta$ parametrises a symmetry of the CFT (when the parameter is switched on uniformly on the whole line).  In particular, provided we properly redefine the operators, all correlation functions should \emph{not depend} on $\theta$. 

For our purposes, it will be enough to use two particularly simple cases. 
For example, the expectation value of the circular Wilson loop does not depend on $\theta$. This means that the vacuum expectation value of the deformed action should be invariant, in other words
      \beq\label{eq:constr0a}
    \langle {\rm P } e^{\int_0^{2 \pi } \delta L(t) }  \rangle_{\oneD} = 1 \,.
    \eeq
    This condition is the simplest way to fix the coupling of the identity operator at order $\mathcal{O}(\mathbf{s}^2)$, as we show in the next section. 

Secondly, we will impose the invariance of certain 2-point functions, as illustrated in Figure \ref{fig:orthoInsert}. We consider 2-point functions involving two identical operators $\Phi_{\perp}^M$, where $M \in \left\{2,3,4,5\right\}$ is a direction orthogonal to $\theta$-deformation. We impose that this 2-point function should not depend on the deformation parameter,
 \beq\label{eq:constr0b}
    \langle {\rm P } e^{\int_0^{2 \pi } \delta L(t) } O_{\Phi_{\perp}^M}(t_1) \, O_{\Phi_{\perp}^M}(t_2) \rangle_{\oneD} =    \langle O_{\Phi_{\perp}^M}(t_1) \, O_{\Phi_{\perp}^M}(t_2) \rangle_{\oneD} = {\rm Pr}(t_1, t_2)  \,.
    \eeq

Studying the latter condition we will deduce a sum rule which constrains all the coefficients $b_{n,2}$, for the operators $O_n$ in our OPE.

We will combine this information with the relation with the cusp anomalous dimension given by the defect configuration of Figure \ref{fig:segment}, and we will see that this leads us to the constraint we are after which is the main result of this paper. For convenience, we list here the key relation we are going to use, transforming the variables of  (\ref{eq:2cusptheta}) to the thermal circle:
\beq\label{eq:constr0c1s}
\langle {\rm P}e^{\int_{t_1}^{t_2} \delta L(t) dt }\rangle_{\oneD} \propto [\epsilon^2 \; P(t_1,t_2)]^{\Gamma_{\rm cusp}(\theta)}\; ,
\eeq
where $\epsilon$ is the UV  cutoff around the points $t_i$, and there is an unfixed, regularisation-dependent proportionality constant. We will see in the next section how we can easily remove such ambiguities and focus on the physical content of the equation. 

As a final comment, let us notice that, in principle, the operators should also be redefined (rotated in R-space) with $\theta$ to ensure invariance of correlation functions. In general, operators would need to be redefined with an expression of the form,
\beq
O_m(t) \rightarrow O_m(t, \theta) = O_m(t) + \sum_{k=1}^{\infty} \mathbf{s}^k \; \sum_n M_{m,n}^{(k)}\; \epsilon^{\Delta_n - \Delta_n} O_n(t),
\eeq
which certain coefficients $M_{m,n}$. The expectation drawn from the ${\cal N}=4$ SYM picture is that for operators taken orthogonal to the deformation, such redefinition is not necessary. 

In the next section we will obtain the constraint originating from this invariance and then impose \eq{eq:constr0c} to obtain the main result for the integrated correlator in closed form, independent on the Wilson coefficients.

\subsection{Extra comments on the operators in the deformed action}

Let us make some additional comments on the operators present in the deformed action. 
The $\mathcal{N}$=4 SYM formalism suggests that  the only operators  present beyond the leading order are $O_{\Phi_{||}}$ and $\mathds{1}$. 
However, in our conformal perturbation theory setup the symmetries of the original theory are somewhat obscured, and 
we cannot definitively exclude that  something more general might happen in our regularisation scheme. Here, we briefly 
discuss what these operators might possibly be and how one could try to fix them. We emphasise, however, that these considerations are not needed for the derivation presented in the next section. 

First, there are certainly some restrictions on the extra operators based on global symmetry considerations. 
 In particular, the $\theta$-rotation on the whole line is a special case of a general symmetry on the conformal manifold, which preserves supersymmetry. We could consider a rotation by angle $\theta$ in a generic plane among the orthogonal directions in R-space, and this should be a natural covariant generalisation of  our deformed action. 
At the leading order, such generic $\theta$-deformation would couple to a tilt operator 
$\delta L = \delta \vec{n}^i  O_{\Phi_{\perp}^i}$, with $\delta n^i$ specifying a direction in R-space, with $\delta n^i \cdot (0,0,0,0,0,1) = 0$ and $|| \delta \vec n || = \mathbf{s}$. At next-to-leading order, the deformed action would need to couple to the two possible tensor structures built with  $\delta \vec{n}^i \delta \vec{n}^j$, indicating that the operators in the action can only be either neutral under R-symmetry, or in the symmetric traceless representation $[2,0]$.

Another natural expectation is that, in any scheme, the only operators which can appear at higher orders in the deformation are the ones which can be built through  the subsequent OPE's of operators at lower orders. 
For example, at next-to-leading order $\mathcal{O}({\bf s}^2)$ we would expect only operators in the OPE $\Phi_{\perp}^1 \Phi_{\perp}^1 \rightarrow  O_n$ \eqref{eq:OPEfusion}. 

Further, the $\theta$-deformation on the whole line obviously preserves conformal symmetry. Therefore, the  beta function for the coupling of every operator in the action should vanish, and remain vanishing as we move $\theta$. This should severely constrain the form of the Wilson coefficients.\footnote{It should in principle fix them completely, apart for possible rearrangements corresponding to reparametrizations of $\theta$ at higher orders. } 
We did not tackle a computation of the beta functions in our scheme. However, we can make some contact with a well-known result of Cardy for the beta functions in conformal perturbation theory with a point-splitting cutoff~\cite{Cardy:1996xt}, which is usually applied to the case of marginal and relevant operators. For perturbations
$ S_{CFT} + \sum_i b_{i} \int d s \,{O}_i(s) $, 
the result of Cardy, specialised to $D=1$, reads
 \beq\label{eq:Cardy}
\beta_{i} = \epsilon \partial_{\epsilon} b_{i} =  (1 - \Delta_i) b_i +  \sum_{k,l} b_k b_l C_{kl}^i + \mathcal{O}(|b|^3),
 \eeq
where $C$'s are the OPE coefficients.
 
 We can see how this result applies to our situation, where the Wilson coefficients $b_i$ additionally scale with $\theta$, $b_i = \sum_k b_{i,k} \mathbf{s}^k$. Consider the case of the marginal operator $O_{\Phi_{\perp}^1}$. Since by assumption $b_{\Phi_{\perp}^1} \equiv \mathbf{s}$ and there is no OPE coefficient $C_{\Phi_{\perp}^1 \Phi_{\perp}^1 \Phi_{\perp}^1 }=0$, (\ref{eq:Cardy}) simply confirms that $\beta_{\Phi_{\perp}^1}$ vanishes up to order $\mathcal{O}(\mathbf{s}^2)$. 
 
 In the case of the identity operator $\mathds{1}$, relation (\ref{eq:Cardy}) expanded up to the same order yields
 \beq
\beta_{\mathds{1}} = b_{\mathds{1},2} \mathbf{s}^2  + (2\mathbb{B} ) C_{\Phi_{\perp}^1 \Phi_{\perp}^1 \mathds{1}} \mathbf{s}^2 + \mathcal{O}(\mathbf{s}^3) , 
 \eeq 
with $C_{\Phi_{\perp}^1 \Phi_{\perp}^1 \mathds{1}} = 1$,  therefore from the vanishing of the beta function we get
 \beq\label{eq:bident0}
  b_{\mathds{1},2} = - 2\mathbb{B} .
 \eeq
This is indeed what we will find, by an independent calculation, in the next section, see \ref{eqn:b02}. 

One would be tempted to apply (\ref{eq:Cardy}) also to fix the couplings for the irrelevant operators. However, this does not seem to be consistent with our results, in particular with the sum rules deduced in the next section. 
We were also not able to reproduce the arguments of (\ref{eq:Cardy}) in application to our specific regularisation scheme. A careful analysis would be needed in order to compute the beta functions for the irrelevant operators in our action, and we will not study this problem here.

In our derivation, we are fixing part of the couplings (or more precisely, sum rules for them\footnote{As we show in the next section, we can compute $b_{\Phi_{||}, 2}$, but only \emph{under the assumption} that the coefficients for other irrelevant operators are zero.}) by computing some observables and imposing that they satisfy the physical invariances of the $\theta$-deformation. We have imposed such conditions only for a couple of special observables. We can imagine that imposing more constraints, in addition to the vanishing of the beta functions, would help fix the Wilson coefficients and clarify whether more operators on top of $\mathds{1}$ and $O_{\Phi_{||}}$ should be included in the action.   
Fixing precisely the couplings might be useful in some applications. In particular, we could use them to compute some physical observables -- for example, one could consider ``multi-cusp'' correlators where different $\theta$'s are switched on on different segments of the line. Clarifying these points could be a fruitful potential direction for future studies, which, however is not critical for the derivation of this paper as we discuss in the next section.

\section{Derivation of the integrated correlator involving the curvature function}\label{sec:derive}

We will use the formal approach described in the previous section to relate the curvature function with an integrated 4-point correlator.
The derivation will be done in two steps: First we deduce a constraint on the Wilson coefficients
$b_{n,k}$ from the requirement that the deformation by $\theta$ applied to the whole space should be a symmetry. More precisely we impose the properties (\ref{eq:constr0a}) and (\ref{eq:constr0b}). Second, we use the relation between the cusp anomalous dimension and the deformation applied to a part of the space (\ref{eq:constr0c}) expanded to the forth order in ${\bf s}$. By using these two equations, we will be able to  derive our final result: the linear combination of the two integrated correlators \eqref{eq:constr1} and \eqref{eq:constr2}, complementary to the one derived in \cite{Drukker:2022pxk} completing the derivation of the two relations found in \cite{Cavaglia:2022qpg}.

 \subsection{Constraining the $b_{\mathds{1},2}$ coupling}
 We begin by considering the constraint \eqref{eq:constr0a} at the ${\rm O}({\bf s}^2)$ order, which gives
\begin{align}\label{invb02}
    {\bf s}^2\,\left[ \int_{\epsilon}^{2\pi} ds_2 \int_{{\rm  max}(0,s_2-2\pi+\epsilon)}^{s_2-\epsilon} ds_1\langle O_{\Phi_\perp^1}(s_1) O_{\Phi_\perp^1}(s_2) \rangle_{\oneD} + \frac{b_{\mathds{1},2}}{\epsilon}\int_{0}^{2\pi} ds\right]  = 0  \;,
\end{align}
where we explicitly written the regularisation of the integrals with the measure $\mu_2$ \eqref{mun} that introduces the cutoff $\epsilon$ to prevent  the coordinates $s_1$ and $s_2$ coming closer than $\epsilon$ to each other. The first term in \eqref{invb02} is the integrated two-point function of the marginal operator $O_{\Phi_{\perp}^1}$ while the second one is the contribution of the identity. Using the definition \eqref{2pt} and the normalisation \eqref{normalisation}, one can solve the integration, which gives
\beq
\int_{\epsilon}^{2\pi} ds_2 \int_{{\rm  max}(0,s_2-2\pi+\epsilon)}^{s_2-\epsilon} ds_1\frac{2{\mathbb B}}{2-2\cos(s_1-s_2)}
=\frac{4\pi{\mathbb B}}{\epsilon}+{\cal O}(\epsilon)\;,
\eeq
thus we obtain
\beq\label{eqn:b02}
    b_{\mathds{1},2} = -2 \;\mathbb{B}\;,
\eeq
which is in agreement with the argument in (\ref{eq:bident0}). 
We will use this relation in the next section.

\subsection{Constraining a combination of $b_{n,2}$ couplings}\label{sec:firstargument}

Now we impose the constraint (\ref{eq:constr0b}).
We take equal indices $N=M$ from the beginning with $M\in\left\{2,3,4,5\right\}$ and expand
to the order $\mathcal{O}({\bf s}^2)$ to obtain for the coefficients of ${\bf s}^2$
\beqa\label{eq:constraint8}
0 &=& \underbrace{\frac{b_{0,2}}{\epsilon} \; \left(\int_0^{2 \pi} ds\, \mu_{1,2}(s;t_1,t_2)
\right)\,  \langle O_{\Phi_{\perp}^M}(t_1)  O_{\Phi_{\perp}^M}(t_2)\rangle_{\oneD} }_{\text{identity contribution }\equiv \;\mathcal{I}_{1-\text{pt}} } \nn\\&&+ \underbrace{\sum_{\Delta_n>1} \epsilon^{ \Delta_n - 1} b_{n,2} \int_{0}^{2 \pi} ds\, \mu_{1,2}(s;t_1,t_2)\;
\langle O_{\Phi_{\perp}^M}(t_1) {O}_{n}(s) O_{\Phi_{\perp}^M}(t_2) \rangle_{\oneD} }_{\text{integrated 3-point }\equiv \;\mathcal{I}_{3-\text{pt}}  }  \\&&+ \underbrace{\int_{0<s_1<s_2<2 \pi} ds_1 ds_2 \; \mu_{2,2}(s_1,s_2;t_1,t_2)\;
\langle O_{\Phi_{\perp}^M}(t_1) O_{\Phi_{\perp}^1}(s_1)  O_{\Phi_{\perp}^1}(s_2) O_{\Phi_{\perp}^M}(t_2) \rangle_{\oneD} }_{\text{integrated 4-point }\equiv \;\mathcal{I}_{4-\text{pt}} }\nn ,
\eeqa
where the integration measure $\mu_{n,m}$ enforce the hard-sphere cutoffs as defined in section~\ref{sec:regularisation}.
The contribution of the first line of (\ref{eq:constraint8}) is given by the trivial integral of the measure $\mu_1$ and it reads
\beq
\mathcal{I}_{1-\text{pt}} = \frac{(2 \pi - 4 {\epsilon})}{\epsilon} \, b_{\mathds{1},2} \, (2 \mathbb{B} ) {\rm P}(t_1,t_2),
\eeq
with $b_{\mathds{1},2} = -2 \mathbb{B}$ given by \eqref{eqn:b02} and another $2 \mathbb{B}$ coming from the special normalisation of the tilt operator $O_{\Phi_{\perp}^M}$. 
The remaining two terms are more involved and they are computed explicitly in appendix \ref{app:integrals1}.
Integrated 3-point functions appearing in the second line can be obtained assuming a generic spectrum with $\Delta_n >1$ at finite coupling. Taking into account the kinematics \eqref{2pt3ptCFT} and normalisation \eqref{cn}, we find that the integral is finite and gives 
\beq\label{eq:first3ptintegrals}
\mathcal{I}_{3-\text{pt}}  = (2 \mathbb{B} ) \,{\rm P}(t_1,t_2)\;\sum_{\Delta_n>1} b_{n,2} \frac{4\, C_n}{\Delta_n-1} \;.
\eeq
The final contribution is given by the 4-point functions in the last line of (\ref{eq:constraint8}). In order to solve this integral it is convenient to use the parametrisation for $G(x)$ given in \eqref{Gwithdelta}. Details of this calculation are described in appendix~\ref{app:integrals} and lead to the following result
\beq\small\begin{split}\label{eq:first4ptintegrals}
\mathcal{I}_{4-\text{pt}}\! = \! (2\mathbb{B} )^2\Biggl(\frac{2 \pi \!-\! 6{\epsilon} }{\epsilon}\!+\! \int_0^{\frac{1}{2} } \!\!dx\! \left[ \frac{\delta G_3(x) }{x^2} \log{x} + \frac{\delta G_1(x) }{x^2} \log{\frac{x^2}{1\!-\!x}} + \frac{\delta G_2(x) }{x^2} \log{\frac{x}{1\!-\!x}}\right]\!\!\Biggl){\rm P}(t_1,t_2),
\end{split}\eeq
where $\delta G_i(x) =G_i(x)-G_{i,\text{tree}}(x)$ with $G_i(x)$ are given in \eqref{G1G2G3} and their tree-level values are
\beq\label{G1G2G3tree}
        G_{1,\text{tree}}(x) = 1\;, \qquad
        G_{2,\text{tree}}(x) = 0\;, \qquad
        G_{3,\text{tree}}(x) = \frac{x^2}{(x-1)^2}\;.
\eeq
Notice that the integration in \eqref{eq:first4ptintegrals} goes in the domain $[0,1/2]$ and, unfortunately, there is no way to extend it to the whole interval $[0,1]$ using crossing \eqref{crossingsym} in a smooth way. This is a manifestation of the scheme dependence of the Wilson coefficients. However, we will see, that in the final result, which only contains physical quantities, the integration can be extended to the whole range naturally.

Once all pieces are combined, from $\mathcal{I}_{1-\text{pt}} +\mathcal{I}_{3-\text{pt}} +\mathcal{I}_{4-\text{pt}}  = 0$ we find
\beqa\label{eq:sumruleconstr}
\!\!\!\!\!\!\sum_{\Delta_n>1}  \! b_{n,2}\frac{C_{n} }{\Delta_n \!-\!1} + \mathbb{B} \!\! \int_0^{\frac{1}{2} }\! \!dx\! \left[ \frac{\delta G_3(x) }{x^2} \log{x} + \frac{\delta G_1(x) }{x^2} \log{\frac{x^2}{1\!-\!x}} + \frac{\delta G_2(x) }{x^2} \log{\frac{x}{1\!-\!x}} \right]\!=\!\mathbb{B} .
\eeqa
This can be seen as a constraint on the weighted sum of the couplings $b_{n,2}$ in terms of the CFT data. At the same time, if we assume that only $O_{\Phi_{||}}$ contributes, the above relation completely fixes the value of the only coefficient $b_{\Phi_{||},2}$! We will use this assumption to analyse $b_{1,2}$ in section~\ref{sec:normphip}. 

Finally, let us rewrite the main result of this section
\eq{eq:sumruleconstr} in terms of the reduced amplitude $\delta f(x)=f(x)-f_{\rm free}(x)$, as defined in \eq{G1G2G3} and \eq{ftree}
\beqa\label{eq:nontribs}
\frac{1}{\mathbb{B}}\!\sum_{\Delta_n>1}   \!b_{n,2}\frac{C_{n} }{\Delta_n -1}  = \!\int_{\delta_x}^{\frac{1}{2} }\! dx \frac{(x\!-\!2) \delta f(x)}{x^3}  - (3 - \mathbb{F} ) \log(\delta_x) + (\mathbb{F}-2) \,\log(2) +  1 .
\eeqa
Note that  the r.h.s. of \eqref{eq:nontribs} is finite in the limit in which the cutoff  $\delta_x\rightarrow 0^+$ since $\delta f(x)\sim \frac{3-\mathbb{F}}{2}x^2$ for $x\rightarrow 0$.

In order to get a closed expression for the integrated correlator, in the next section we make another calculation involving deformation only on a part of the thermal circle of the 1D CFT.

\subsection{Matching with the cusp anomalous dimension}\label{sec:secondder}
So far we obtained the nontrivial equation (\ref{eq:nontribs}) involving the Wilson coefficients $b_{n,2}$ relating them to the OPE coefficients $C_n$ and ${\mathbb B}$ and to an integral of the reduced correlator $f(x)$. 
In this section we exploit the cusp deformation introduced in section \ref{sec:deformations} to conclude our derivation. We start from the equation ~\eqref{eq:constr0c1s}, which can be written in the form
\beq\label{eq:constr0c1}
\langle {\rm P}e^{\int_{t_1}^{t_2} \delta L(t) dt }\rangle_{\oneD}=K\; [P(t_1,t_2)]^{\Gamma_{\rm cusp}(\theta)}\; ,
\eeq
where $K$ is a non-physical renormalisation constant (e.g. depending on the UV cutoffs), while $\Gamma_{\text{cusp}}(\theta)$, independent on the scheme, is the cusp anomalous dimension discussed in section \ref{sec:deformations}. 
In order to get rid of the constant $K$ we consider the following expression
\beq
\label{eq:general}\l\frac{\partial_{t_1} \partial_{t_2} \log{\langle {\rm P}e^{\int_{t_1}^{t_2} \delta L(t) dt }\rangle_{\oneD} } }{{\rm P}(t_1,t_2)}= -2 \Gamma_{\text{cusp}}(\theta)\;,
\eeq
as follows from \eq{eq:constr0c1}. Notice that now both l.h.s. and r.h.s. are finite quantities.
Next we expand both sides in powers of ${\bf s}$.
It is useful to denote the expansion coefficients as follows
\beq\label{ABexp}
\langle {\rm P}e^{\int_{t_1}^{t_2} \delta L(t) dt }\rangle_{\oneD}= 1
    +  A (t_1,t_2)\,{\bf s}^2
    +  B (t_1,t_2)\,{\bf s}^4 + \mathcal{O}({\bf s}^6) ,
 \eeq
where $A(t_1,t_2)$ and $B(t_1,t_2)$ are defined explicitly in appendix \ref{app:cuspexp} and contain contributions from various integrated correlation functions,  schematically
\beq
A(t_1,t_2) = (\texttt{1-pt}) + (\texttt{2-pt}) , \;\;\;B(t_1,t_2) = (\texttt{1-pt}) + (\texttt{2-pt}) + (\texttt{3-pt}) + (\texttt{4-pt})\,.
\eeq
Those terms can also be seen from \eq{eq:expansion}. The only difference is that we allow now for multiple $O_n$ and not only $\Phi_{||}$.

Even though we only have the action \eq{eqn:deltaLdef} at the order ${\bf s}^2$, it is easy to convince ourselves that only the identity operator at order ${\bf s}^4$ can contribute (which we do take into account), but nothing else which could appear in the action at higher orders. The function $A(t_1,t_2)$ takes contributions from the integrated 1-point functions (of the identity operator), as well as 2-point function of the line-deformation operator $O_{\Phi_{\perp}^1}$, while $B(t_1,t_2)$ contains contributions from: integrated 1-point function of the identity operator at the next order, integrated 2-point functions of generic operators in the action, integrated 3-point functions involving two $O_{\Phi_{\perp}^1}$ and a third generic operator and integrated 4-point correlators of the line-deformation scalars. 
From the above expansion it follows
\begin{align}\label{logW}
 \log\,\langle {\rm P}e^{\int_{t_1}^{t_2} \delta L(t) dt }\rangle_{\oneD} =  A (t_1,t_2)\,{\bf s}^2 + \bigg[B (t_1,t_2) - \frac{A^2 (t_1,t_2)}{2}\bigg]\,{\bf s}^4 + \mathcal{ O}({\bf s}^6) \;.
\end{align}
In the following we compute the two combinations appearing in \eqref{logW} in terms of the Bremsstrahlung \eqref{bremdef} and Curvature functions \eqref{curvaturedef} arising from the expansion of the cusp anomalous dimension.

\paragraph{First order. }
At leading order in ${\bf s}$, (\ref{eq:general}) becomes 
\beq\label{eqn:SegmentFirstOrder}
\frac{\partial_{t_1} \partial_{t_2} A(t_1,t_2)  }{{\rm P}(t_1,t_2)} = -2 \mathbb{B} \,.
\eeq
As from (\ref{eq:simpleA}) we have
\beq
\partial_{t_1} \partial_{t_2} A(t_1,t_2)  =
-\langle O_{\Phi_{\perp}^1}(t_1)O_{\Phi_{\perp}^1}(t_2)\rangle_{\oneD}\,,
\eeq
the equation \eq{eqn:SegmentFirstOrder} does indeed hold as a consequence of the 2-point function normalisation \eqref{normalisation}. 

\paragraph{Next-to-leading order. }
The next-to-leading order constraint deriving from (\ref{eq:general}) reads
\beq\label{eq:NLOconstraint}
\frac{\partial_{t_1} \partial_{t_2} \left(B(t_1,t_2)  - \frac{A^2 (t_1,t_2)}{2}\ \right) }{{\rm P}(t_1,t_2)} = - \frac{\mathbb{B} + \mathbb{C} }{2} .
\eeq 
The evaluation of the l.h.s. in our regularisation scheme is rather long, and is spelled out in appendices (the final result is deduced in (\ref{eq:finalABresult}), relying on explicit calculations which are stored  in appendix \ref{app:integrals}).   The main nontrivial contributions, as in the calculation of the previous section,  come from the integrated 3-point functions of the type $\langle  O_{\Phi_{\perp}^1} O_{\Phi_{\perp}^1} O_{n} \rangle_{\oneD}$ as well as from the integrated 4-point function $ \langle O_{\Phi_{\perp}^1}O_{\Phi_{\perp}^1}O_{\Phi_{\perp}^1}O_{\Phi_{\perp}^1} \rangle_{\oneD}$. After the dust settles,  the divergences cancel and  (\ref{eq:NLOconstraint}) becomes
\beq\begin{split}\label{eq:prelast}
-\frac{1}{2} \left( \mathbb{C} + \mathbb{B} \right)  &= - (2 \mathbb{B} ) \sum_{\Delta_n>1} \;  b_{n,2}\;\frac{ 4 C_{n} }{\Delta_n -1} - (2 \mathbb{B})^2 (2 - \mathbb{F} )\,\left( 1 + \log 4 \right) + 4  \;\mathbb{B}^2  \\
&- (2 \mathbb{B} )^2 \,\left[ \int_{\delta_x}^{\frac{1}{2}} dx \frac{(2 x-3) ((x-1) x+1) \delta f(x)}{(x-1) x^3}\, dx + \frac{3}{2} (3 - \mathbb{F} ) \log\delta_x \right].
\end{split}\eeq
Notice that the sum, containing the Wilson coefficients, appearing in \eqref{eq:prelast} is exactly the same as in \eq{eq:nontribs}. Thus we can exclude them completely obtaining a closed expression for the integrated correlator.

\paragraph{A new linear combination of integrated correlators.}
Substituting in \eqref{eq:prelast} the sum over the Wilson coefficients $b_{n,2}$ given by (\ref{eq:nontribs}), we obtain
\beqa\label{eq:curvconstrfinal}
- \frac{\mathbb{C} + \mathbb{B}}{8\,  \mathbb{B}^2} &=& - 3 + \mathbb{F} +\frac{1}{2} (3 - \mathbb{F} ) \log{\delta_x} + \int_{\delta_x}^{\frac{1}{2} } dx \; \delta f(x) \left( \frac{1}{x^3}-\frac{3}{x}+\frac{1}{x-1} \right) ,
\eeqa
which is a constraint involving only CFT data. This is the main result of our derivation. 
It is simple to verify that \eqref{eq:curvconstrfinal} is a linear combination of the two constraints (\ref{eq:constr1}) and (\ref{eq:constr2}) originally found in \cite{Cavaglia:2022qpg}. Indeed, using crossing symmetry \eqref{crossingsym} to rearrange the integration domain, they can be rewritten as follows
\begin{align}
 \label{const1app} \texttt{Constraint1:}&     \int_{\delta_x}^{\frac{1}{2}} \!\delta f(x)  \!\left( \frac{1}{x^3}\!-\!\frac{2}{x^2}\!+\!\frac{1}{x}\!+\!\frac{1}{x\!-\!1
   }\!\right)\!dx +\frac{ 3 \!- \!\mathbb{F} }{2} \log{\delta_x} \!+\! 3\! - \!\mathbb{F} = \frac{3\mathbb{C}\!-\!\mathbb{B}}{8 \mathbb{B}^2},\\
 \texttt{Constraint2:}& \int_{0}^{\frac{1}{2}}\! \delta f(x)\left(\frac{1}{x} -\frac{1-x}{x^2} \right)dx = \frac{\mathbb{C}}{4\mathbb{B}^2} + \mathbb{F}-3.\label{const2app} 
\end{align}
Combining \eqref{const1app} and \eqref{const2app}  with coefficients $1 \times \texttt{Constraint1} - 2 \times \texttt{Constraint2}$, it is simple to verify that (\ref{eq:curvconstrfinal}) is perfectly reproduced. This, together with the combination $-2 \times \texttt{Constraint1} + 3 \times \texttt{Constraint2}$ \eqref{Nadav} derived in \cite{Drukker:2022pxk} we conclude the proof of those relations.

\subsection{Normalisation of $\Phi_{||}$}\label{sec:normphip}

Assuming that only the operator $O_{\Phi_{||}}$ contributes to the sum in \eqref{eq:deltaA}, it is possible to link our conformal perturbation theory setup with the usual ${\cal N}=4$ SYM expression for the MWL computing the only remaining Wilson coefficient $b_{1,2} \equiv b_{\Phi_{||} ,2}$. In order to evaluate it, we can exploit the constraint \eqref{eq:nontribs}.
Using it we can evaluate 
$b_{1,2}$ analytically  both at weak and strong coupling, from the known expression for $f(x)$~\cite{Kiryu:2018phb,Cavaglia:2022qpg,Ferrero:2021bsb}.

\paragraph{Weak coupling.}
At weak coupling, the integral of $f(x)$ has to be treated carefully. Indeed, it contains an anomalous term given by the divergence of the superconformal block of the long-multiplet at $x\sim 0$ for $\Delta_1=1$. Following the same logic of \cite{Cavaglia:2022qpg}, we can quantify the ``anomaly'' term to be $2C_1^2/(\Delta_1-1)^2$ (in the terminology of \cite{Cavaglia:2022qpg}). Details of the computation are given in appendix \ref{app:appD}. Using the regularisation given in \eqref{fint1}, the integral appearing in \eqref{eq:nontribs} is computed using the representation of $f(x)$ in terms of harmonic polylogarithms (HPL) implemented in the Mathematica package \cite{Maitre:2005uu}. Using the HPL's properties, the integral can be solved recursively using integration by parts. Plugging it in the constraint  \eqref{eq:nontribs} and solving for $b_{1,2}$ we have
\begin{align}\label{b12weak}
b_{1,2}&=\frac{g}{\sqrt{2}}-\frac{g^3\left(\pi ^2-6\right) }{3 \sqrt{2}}+\frac{g^5 \left(-108 (\zeta_3+3)+12 \pi ^2+5 \pi ^4+576 \log
  2\right)}{18 \sqrt{2}}+...\,,
\end{align}
where $...$ stand for higher orders in the coupling. An additional order is presented in appendix \ref{app:appD}.
Notice that, the sign of the result \eqref{b12weak} depends on the choice of the one of the square root of $C_1^2$, while the sign of the product $b_{1,2}\,C_1$ is fixed. We chose the positive sign for the square root.

\paragraph{Strong coupling.}
In this regime the integral in \eqref{eq:nontribs} does not have any additional divergences. Strong coupling data are given in \cite{Ferrero:2021bsb}. Similarly to the weak coupling case, the integral is computed using the representation of $f(x)$ in terms of Harmonic polylogarithms. The integral is divergent for $\delta_x\rightarrow 0$, but all the divergences are nicely cancelled by the $(3-\mathbb{F})\log \delta_x$ regulator as expected. We obtain
\begin{align}
b_{1,2}\!=\!\frac{\!1\!\!+\!\log 2}{\sqrt{2/5}\,\pi }g\!-\!\frac{ \!\!149\!-\!8 \pi ^2\!+\!17 \log 2}{96\,\sqrt{2/5}\, \pi
   ^2}\!+\!\frac{ \!\!5616 \zeta_3\!+\!21003\!-\!1232 \pi ^2\!+\!45 \log 2 (256 \log 2\!-\!129)}{18432\,\sqrt{2/5}\, \pi ^3
   g}\!+\!...
\end{align}
where $...$ stand for higher orders in $1/g$. Two additional orders are included in appendix \ref{app:appD}.

\paragraph{Comparison with field normalisation.} Now comparing \eq{eq:action2} and \eq{eqn:deltaLdef} at the order ${\bf s}^2$ we conclude that we should impose, in the case when only one marginal operator contributes
\beq
\Phi_{||} = -2 b_{1,2}\epsilon^{\Delta_1-1}O_{\Phi_{||}}\,.
\eeq
The operator $O_{\Phi_{||}}$ has the standard unit normalisation in our conventions, from where we can conclude that the scalar of ${\cal N}=4$ SYM has to be normalised as follows
\beq
\langle \langle \Phi_{||}(t_1) \;\Phi_{||}(t_2) \rangle\rangle
=4 b_{1,2}^2 \epsilon^{2\Delta_1-2}\frac{ \delta_{ij}}{x_{12}^{2 \Delta_i}}\;.
\eeq
As from above we know $b_{1,2}$, this would then give us the normalisation of the non-protected scalar $\Phi_{||}$ in our regularisation scheme. We can now quickly test this relation at weak coupling: at the leading order there should not be difference between 
normalisation of $\Phi_{||}$ and $\Phi_{\perp}$
as at tree level there is no interaction with the MWL, thus $4 b_{1,2}^2 \epsilon^{2\Delta_1-2}$ should coincide with $2{\mathbb B}\simeq 2g^2$ at the leading order, which is indeed the case as one case see from \eq{b12weak}.
Furthermore, we notice that $\sqrt{{\mathbb B}/2}\simeq \frac{g}{\sqrt{2}}-\frac{\pi ^2 g^3}{3 \sqrt{2}}+\frac{5 \pi ^4 g^5}{18 \sqrt{2}}+O\left(g^6\right)$ reproduces all terms in $b_{1,2}^2$ with maximal power of $\pi$, but otherwise there is no reason for the scalars to have the same normalisation beyond the leading order, as one is protected and  the other is not in the interacting case.

\section{Discussion}
In this paper, we completed the proof of the integral constraints presented in~\cite{Cavaglia:2022qpg}. A linear combination of those relations was already derived in~\cite{Drukker:2022pxk} using a geometrical approach. Here, exploiting the invariance of the defect CFT under R-space rotation and relating it to the generalised cusp anomalous dimension, we provided the derivation of second relation, thereby completing the proof.

A possible future direction is to derive
constraints on multi-point correlation functions e.g. 6-point functions, which should be related to the higher orders in expansion of the generalised cusp anomalous dimension with respect to the $\theta$ and $\phi$ deformations. Finding a shortcut method for deriving such relations between the correlation functions and the spectrum of the deformations would generate, in principle, an  infinite amount of additional constraints on the 1D CFT, which could be sufficient for its complete solution.
We also expect that at each order there should be an increasing  number of such constraints -- like we found $2$ of them for $4$-point function, one could speculate that there should be at least $3$ non-trivial constraints for $6$-point functions. 
We also note that the data coming from the integrability side become richer with each order in $\theta$ or $\phi$: the Bremsstrahlung function contains only powers of $\pi$ in its perturbation theory, the Curvature function already brings in zeta functions, and we expect MZV's to come from the next order as well. The higher point correlation functions in the current context were studied recently in \cite{Barrat:2021tpn,Barrat:2022eim} at weak coupling, which could give a starting point in this investigation. 

Furthermore, one can use the integrability data for the cusp with arbitrary operators sitting at the cusp, to constrain more complicated correlators with non-BPS external legs.

Another interesting direction is to extend our derivation to other integrable gauge theories. The most natural choice would be the three-dimensional ABJM theory where 1D superconformal defect theories supported by Wilson loops were defined in \cite{Bianchi:2017ozk,Bianchi:2018scb} and recently studied in \cite{Bianchi:2020hsz,Gorini:2022jws}. The cusp anomalous dimension was studied in \cite{Griguolo:2012iq,Bonini:2016fnc} and the Bremsstrahlung function is known exactly \cite{Bianchi:2014laa,Correa:2014aga,Bianchi:2017svd,Bianchi:2018scb} (see also \cite{Drukker:2019bev}). The curvature function is still not known since an integrability formulation for the cusped Wilson line is still lacking.\footnote{However, based on the $\mathcal{N}$=4 SYM example it seems very likely that this could be obtained by deforming the QSC  for local operators in ABJM theory. The latter is known~\cite{Cavaglia:2014exa,Bombardelli:2017vhk}. } However, all the conformal perturbation theory approach we have described seems ready for application  in this context, and also the geometrical  argument of \cite{Drukker:2022pxk} should lead to a new non-trivial constraint for the line defect in ABJM theory. 

It would be also interesting to investigate the fishnet limit~\cite{Erickson:1999qv,Zamolodchikov:1980mb,Gurdogan:2015csr,Caetano:2016ydc,Kazakov:2018gcy,Pittelli:2019ceq,Gromov:2021ahm} and integrated operators in this simpler theory, where one can hope to advance analytically more easily and in particular to get the answer to the fundamental question if integrability for the spectrum in combination with conformal symmetry is sufficient to solve this type of beyond-the-spectrum observables.

\acknowledgments
We thank N.~Drukker, M.~Gaberdiel, C.~Herzog, Z.~Komargodski, G.~Korchemsky, Z. Kong, R.~Tateo and G.~Watts for discussions, and especially Amit Sever for sharing expertise on similar techniques  used in another context in  \cite{Gabai:2022vri}. AC thanks Marco Baggio for lectures and very inspiring  discussions on conformal perturbation theory.
The work of AC, NG and MP is
supported by European Research Council (ERC) under
the European Union’s Horizon 2020 research and innovation programme (grant agreement No. 865075) EXACTC. NG is also partially supported by the STFC grant (ST/P000258/1).

\appendix

\section{The curvature function}\label{app:curvature}

The Curvature function was computed by means of QSC in terms of a double contour integral in ~\cite{Gromov:2015dfa}. 
In the $\phi\rightarrow 0$ limit, it reduces to \eqref{curvaturedef} where the kernel is is given by
\beq
\label{Gamma0}
K_0(u)=\d_u\log\frac{\Gamma(i u+1)}{\Gamma(-i u+1)}\,,
\eeq
and the integrand $F$ is
\beqa\label{Fxy}
	F[x,y]&&=-\frac{8 i \sinh \left(2 \pi  u_x\right) u_x  u_y x^2S_0(y) }{I_1(4 g \pi ){}^2}
	\\ \nn
	&&+S_0(y){}^2\left[\frac{8 i x y I_2(4 g \pi ) u_x u_y}{g
   \pi  \left(x^2-1\right) I_1(4 g \pi ){}^3}-\frac{8 i x y I_2(4 g \pi ) u_x u_y}{g \pi
   \left(y^2-1\right) I_1(4 g \pi ){}^3}+\frac{32 i x y u_x u_y}{I_1(4 g \pi ){}^2}\right]
	\\ \nn
	&&+\sinh ^2\left(2 \pi  u_y\right) \left[\frac{4 i x y I_2(4 g \pi ) u_x u_y}{g
   \pi  \left(x^2-1\right) I_1(4 g \pi ){}^3}+\frac{16 i x y u_x u_y}{I_1(4 g \pi
   ){}^2}\right]
	\\ \nn
	&&+\sinh \left(2 \pi  u_y\right) \left[\frac{4 i x u_x u_y
   y^2}{\left(x^2-1\right) I_1(4 g \pi )}
	-\frac{8 i x \sinh \left(2 \pi  u\right) u_x u_y
   y}{I_1(4 g \pi ){}^2}
	-\frac{8 i u_x u_y S_1(x) y}{g I_1(4 g \pi ){}^2}
	\right.
	\\ \nn
	&& \left.
	-\frac{16 i x u_x 
   u_y}{\left(y^2-1\right) I_1(4 g \pi )}+\left(-\frac{8 i x y I_2(4 g \pi ) u_x u_y}{g \pi
   \left(x^2-1\right) I_1(4 g \pi ){}^3}-\frac{32 i x y u_x u_y}{I_1(4 g \pi ){}^2}\right)
   S_0(y)\right]
	\\ \nn
	&&+S_1(y)\left[\frac{8 i x y u_x u_y}{g \left(x^2-1\right) I_1(4 g \pi )}-\frac{8 i x
   y u_x u_y}{g \left(y^2-1\right) I_1(4 g \pi )}\right]
	\\ \nn
	&&
	+S_0(x) \left[S_0(y)\left(\frac{16 i
   u_x u_y}{I_1(4 g \pi ){}^2}-\frac{16 i y^2 u_x u_y}{I_1(4 g \pi ){}^2}\right)
   -\frac{4 i x I_2(4 g \pi ) u_x u_y S_1(y)}{g^2 \pi  \left(x^2-1\right) I_1(4 g \pi
   ){}^3}\right]
	\\ \nn
	&+&S_0(y) \left[\frac{8 i x u_x u_y y^2}{\left(x^2-1\right) I_1(4 g \pi
   )}
	+\frac{8 i x u_x u_y}{I_1(4 g \pi )}
	-\frac{8 i x u_x u_y}{\left(x^2-1\right) I_1(4 g \pi
   )}
	+\frac{32 i x u_x u_y}{\left(y^2-1\right) I_1(4 g \pi )}
	\right. \\ \nn
	&&\left. +S_1(x)\left(-\frac{4 i x I_2(4 g \pi
   ) u_x u_y}{g^2 \pi  \left(x^2-1\right) I_1(4 g \pi ){}^3}-\frac{16 i x u_x u_y}{g I_1(4 g
   \pi ){}^2}\right)
	\right. \\ \nn
	&& \left.
	+S_1(y)\left(\frac{4 i x I_2(4 g \pi ) u_x u_y}{g^2 \pi
   \left(x^2-1\right) I_1(4 g \pi ){}^3}+\frac{16 i x u_x u_y}{g I_1(4 g \pi ){}^2}\right)
   \right],
\eeqa
with
\beq
	S_0(x)=\sum\limits_{n=1}^\infty \frac{I_{2n+1}(4\pi g)}{x^{2n+1}},\ \ \
	S_1(x)=\sum\limits_{n=1}^\infty \frac{2nI_{2n}(4\pi g)}{\pi x^{2n}}\;.
\eeq
These are functions of $x$, $y$ that are related to $u_x$, $u_y$ by the usual Zhukovsky map
\beq\label{zhu}
	x+\frac{1}{x}=\frac{u_x}{g},\ \ \ |x|\geq 1\ ,
\eeq
(same for $y$, $u_y$), which resolves the cut $[-2g,2g]$ around which the integrals in \eqref{curvaturedef} run. 
We attach with this paper, a \texttt{Mathematica} notebook ``\texttt{Curvature.nb}'' which computes the curvature function numerically for a given value of the coupling $g$.

\section{Expansion of the segment}\label{app:cuspexp}

The quantities $A$, $B$ appearing in \eqref{ABexp} are defined through the following expansion
\begin{equation}\begin{split}
\bigg\langle\; \text{P\,exp}\biggl[\int_{t_1}^{t_2} dt\,\biggl( {\bf s} \; O_{\Phi_{\perp}^1}(t) + \sum_{\Delta_n >1} &\epsilon^{\Delta_{n}-1}\,O_n(t) \, \sum_{k=2}^{\infty} b_{n,k} \,{\bf s}^{k}\, \biggl) \biggl]\; \bigg\rangle_\oneD \\
&\equiv 1 + {\bf s}^2 \; A(t_1, t_2) + {\bf s}^4 \; B(t_1, t_2) + \dots .
\end{split}
\end{equation}
The leading order receives contributions from integrated 1- and 2-point functions such that
\beq
    A(t_1,t_2) = K_{\mathds{1}}^{\text{1-pt}} +  K^{\text{2-pt}} ,
    \eeq
with the explicit integrals given by
\beqa
K_{\mathds{1}}^{\text{1-pt}} &\equiv& \frac{b_{0,2} }{\epsilon}\int_{t_1}^{t_2} ds \;\mu_{1,2}\; \langle \mathds{1}(s) \rangle_\oneD ,\\
K^{\text{2-pt}}
&\equiv& \int_{t_1 < s_1 < s_2 <t_2}  ds_1\, ds_2 \; \mu_{2,2} \; \,\langle O_{\Phi_\perp^1} (s_1) O_{\Phi_\perp^1} (s_2) \rangle_\oneD\; ,
\eeqa
where the integration measures $\mu_{n,m}$ enforcing the cutoff is given in (\ref{munm}). 

\noindent
The next-to-leading order $B(t_1, t_2)$ is defined by the sum 
\beq
    B(t_1,t_2) = L_{\mathds{1}}^{\text{1-pt}} + L_{\mathds{1}}^{\text{2-pt}} + L_{\mathds{1}}^{\text{3-pt}}+\sum_{\Delta_n>1} \left( L_{O_n}^{\text{2-pt}} + L_{O_n}^{\text{3-pt}} \right) + L^{\text{4-pt}} .
\eeq
The contributions involving the identity are given by
\begin{align}
L_{\mathds{1}}^{\text{1-pt}} &\equiv \frac{b_{0,4} }{\epsilon}\int_{t_1}^{t_2} ds \;\mu_{1,2}\; \langle \mathds{1}(s) \rangle_\oneD\; ,\\
L_{\mathds{1}}^{\text{2-pt}}
&\equiv \frac{b_{0,2}^2}{\epsilon^2}\; \int_{t_1 < s_1 < s_2 <t_2}  ds_1\, ds_2 \; \mu_{2,2}  \;\langle \mathds{1}(s_1)\mathds{1}(s_2) \rangle_\oneD\; , \\
L_{\mathds{1}}^{\text{3-pt}}
&\equiv \frac{b_{0,2}}{\epsilon}\; \int_{t_1 < s_1 < s_2 < s_3 <t_2}  ds_1\, ds_2 \,ds_3\; \mu_{3,2}  \;{\Big [ }\langle \mathds{1}(s_1) O_{\Phi^1_{\perp}}(s_2) O_{\Phi^1_{\perp}}(s_3)  \rangle_\oneD \\
& \qquad\qquad\qquad\qquad\qquad+ \langle  O_{\Phi^1_{\perp}}(s_1) \mathds{1}(s_2) O_{\Phi^1_{\perp}}(s_3)  \rangle_\oneD  + \langle  O_{\Phi^1_{\perp}}(s_1) O_{\Phi^1_{\perp}}(s_2)  \mathds{1}(s_3) \rangle_\oneD {\Big ] }\;.\nn
\end{align}
The contributions involving $O_n$ are given by
\begin{align}
L_{O_n}^{\text{2-pt}}
&\equiv {b^2_{n,2}}\,{\epsilon^{2\Delta_n-2} }\; \int_{t_1 < s_1 < s_2 <t_2}  ds_1\, ds_2 \; \mu_{2,2}  \;\langle O_n(s_1) O_n(s_2) \rangle_\oneD \;,\\
L_{O_n}^{\text{3-pt}}
&\equiv {b_{n,2}}\,{\epsilon^{\Delta_n-1} }\; \int_{t_1 < s_1 < s_2 < s_3 <t_2}  ds_1\, ds_2 \,ds_3\; \mu_{3,2}  \;{\Big [ }\langle {O_n}(s_1) O_{\Phi^1_{\perp}}(s_2) O_{\Phi^1_{\perp}}(s_3)  \rangle_\oneD \\
&\qquad\qquad\qquad\qquad + \langle  O_{\Phi^1_{\perp}}(s_1) {O_n}(s_2) O_{\Phi^1_{\perp}}(s_3)  \rangle_\oneD  + \langle  O_{\Phi^1_{\perp}}(s_1) O_{\Phi^1_{\perp}}(s_2)  O_n(s_3) \rangle_\oneD  {\Big ] }\;.\nn
\end{align}
Finally, the $4-$point contribution of $O_{\Phi_\perp^1}$ is
\begin{align}
L^{\text{4-pt}}
&\equiv \!\!\int_{t_1 < s_1 < s_2 < s_3 <s_4 <t_2} \!\!\!\!\!\!\!\!\!\!\!\!ds_1\, ds_2 \,ds_3\,ds_4\; \mu_{4,2} \;\langle  O_{\Phi^1_{\perp}}(s_1) O_{\Phi^1_{\perp}}(s_2) 
O_{\Phi^1_{\perp}}(s_3) O_{\Phi^1_{\perp}}(s_4)  \rangle_\oneD\; . \label{eqn:defL4pt}
\end{align}
Rather than evaluating this contribution directly, we note that what enters our derivation is the expression differentiated w.r.t. the endpoints $t_1$, $t_2$. These parameters enter the expression as integration limits, and differentiating removes two integrations. We then find
\begin{align}\label{eq:simpleA}
    \partial_{t_2}\partial_{t_1}A(t_1,t_2) &= - \langle O_{\Phi_\perp^1} (t_1+\epsilon) O_{\Phi_\perp^1} (t_2-\epsilon) \rangle_\oneD\nn \\
    &= - 2 \mathbb{B} \, {\rm P}(t_1+{\epsilon},t_2-{\epsilon}) =  - 2 \mathbb{B} \, {\rm P}(t_1,t_2) + \mathcal{O}({\epsilon}).
    \;
\end{align}
Plugging this expression into the first order expansion~\eqref{eqn:SegmentFirstOrder}, we see that we get a match.
Next, we need the two combinations in~\eqref{eq:NLOconstraint}. 
Let us start with the term involving $A(t_1,t_2)$. 
We get
\begin{equation}\label{eq:Ksq} 
    \frac{1}{2}\partial_{t_2}\partial_{t_1}A^2(t_1,t_2) =
    \frac{1}{2}\partial_{t_2}\partial_{t_1} \bigg[K_{\mathds{1}}^{\texttt{1-pt}}\bigg]^2
    +\frac{1}{2} \partial_{t_2}\partial_{t_1} \bigg[K^{\texttt{2-pt}} \bigg]^2
    +\partial_{t_2}\partial_{t_1}\bigg[K_{\mathds{1}}^{\texttt{1-pt}}\times K^{\texttt{2-pt}}  \bigg]
    \;,
\end{equation}
where the differentiated kernels are given by 
\begin{align}
    \partial_{t_2}\partial_{t_1} \bigg[K_{\mathds{1}}^{\texttt{1-pt}}\bigg]^2 &= -2\, \frac{b_{0,2}^2}{\epsilon^2} \;, \\
    \label{eqn:pink1}
    \partial_{t_2}\partial_{t_1}
    \bigg[K^{\texttt{2-pt}} \bigg]^2 &= 
    \partial_{t_2}\partial_{t_1}
    \bigg[
    \int_{t_1 + \epsilon}^{t_2-2\epsilon} ds_1
    \int_{s_1 + \epsilon}^{t_2 - \epsilon}  ds_2 \, \langle O_{\Phi^1_\perp}(s_1) O_{\Phi^1_\perp}(s_2) \rangle_\oneD
    \bigg]^2
    \;,\\
    \label{eqn:K1ptK2pt}\partial_{t_2}\partial_{t_1}\bigg[K_{\mathds{1}}^{\texttt{1-pt}}\times K^{\texttt{2-pt}}  \bigg] &= -\frac{b_{0,2}}{\epsilon}
    \int_{t_1 + \epsilon}^{t_2 - 2\epsilon} ds_1\, \langle O_{\Phi_\perp^1}(s_1)O_{\Phi_\perp^1}(t_2 - \epsilon) \rangle_\oneD
    \\
    &\!\!\!\!\!\!\!\!\!\!\!\!\!\!\!\!\!\!\!\!\!\!\!\!\!\!\!\!\!\!\!\!\!\!\!\!\!\!\!\!\!\!\!\!\!\!\!\!\!\!\!\!\!\!\!\!\!\!\!- \frac{b_{0,2}}{\epsilon} 
    \int_{t_1 + 2 \epsilon}^{t_2 - \epsilon} ds_2\, \langle O_{\Phi_\perp^1}(t_1 + \epsilon)O_{\Phi_\perp^1}(s_2) \rangle_\oneD
    - \frac{b_{0,2}}{\epsilon} (t_2 - t_1 - 2\epsilon) \,
    \langle O_{\Phi_\perp^1}(t_1 + \epsilon)  O_{\Phi_\perp^1} (t_2 - \epsilon) \rangle_\oneD\;.\nn
\end{align}
The integral~\eqref{eqn:pink1} will be evaluated in combination with the four-point function~\eqref{eqn:pink2}. On the other hand, integral
\eqref{eqn:K1ptK2pt} can be computed easily by itself. Indeed, Taylor expanding in $\epsilon$ and discarding terms $\mathcal{ O}(\epsilon)$, we get 
\begin{multline}\label{eqn:2ptSqMixed}
    \partial_{t_2}\partial_{t_1}\bigg[K_{\mathds{1}}^{\texttt{1-pt}}\times K^{\texttt{2-pt}}  \bigg] =
    -\frac{4\,\mathbb{B}\,b_{0,2}}{\epsilon^2}     +\frac{\mathbb{B}\,b_{0,2}}{3}
    \\
    -\frac{2\,\mathbb{B}\,b_{0,2}}{\epsilon}\bigg[\Big(t_2 - t_1 - 2\epsilon - 2\,\sin(t_2 - t_1 - 2 \epsilon)\Big) {\rm P}(t_1+{\epsilon},t_2-{\epsilon})\bigg] + \mathcal{O}(\epsilon)
    \;.
\end{multline}
Next, we evaluate the derivative with respect to $t_1$ and $t_2$ of $B(t_1, t_2)$. 
We display the result term by term. Let's consider first the contributions involving the identity operator. Integrated one- and two-point functions are simply given by
\begin{align}
    \partial_{t_1}\partial_{t_2} L_{\mathds{1}}^{\text{1-pt}} 
    = 0 \;,\qquad
    \partial_{t_1}\partial_{t_2} L_{\mathds{1}}^{\text{2-pt}}
    = -\frac{b_{0,2}^2}{\epsilon^2} \;,
\end{align}
while the three-point one is
\begin{multline}\label{eqn:3ptIdSegment}
    \partial_{t_1}\partial_{t_2} L_{\mathds{1}}^{\text{3-pt}} 
    = 
    -\frac{b_{0,2}}{\epsilon}\int_{t_1  +2 \epsilon}^{t_2 - 2\epsilon}ds_2
    \bigg[
    \langle O_{\Phi_\perp^1}(t_1 + \epsilon)O_{\Phi_\perp^1}(s_2)\mathds{1}(t_2 -  \epsilon)    \rangle_\oneD 
    \\ 
    +
    \langle O_{\Phi_\perp^1}(t_1 + \epsilon)\mathds{1}(s_2)O_{\Phi_\perp^1}(t_2 - \epsilon)    \rangle_\oneD
    +
    \langle \mathds{1}(t_1 + \epsilon)O_{\Phi_\perp^1}(s_2)O_{\Phi_\perp^1}(t_2 - \epsilon)    \rangle_\oneD
    \bigg]\;.
\end{multline}
Integrating~\eqref{eqn:3ptIdSegment}, Taylor expanding in $\epsilon$, and discarding terms ${\rm O}(\epsilon)$, we get
\begin{multline}\label{eqn:3ptPhiSegmentEps}
    \partial_{t_1}\partial_{t_2} L_{\mathds{1}}^{\text{3-pt}} 
    =
    -\frac{4\,\mathbb{B}\,b_{0,2}}{\epsilon^2}+ \bigg[8\,\mathbb{B}\,b_{0,2}\,{\rm P}(t_1+ {\epsilon},t_2- {\epsilon}) + \frac{\mathbb{B}\,b_{0,2}}{3}\bigg]  \\
    -\frac{2\,\mathbb{B}\,b_{0,2}}{\epsilon}\bigg[\Big(t_2 - t_1 - 2\epsilon - 2\sin(t_2 - t_1 - 2 \epsilon)\Big) {\rm P}(t_1+{\epsilon},t_2-{\epsilon})\bigg]
    + \mathcal{O}(\epsilon)
    \;.
\end{multline}
Secondly, we focus on the terms involving operators in the long multiplet with dimension $\Delta_n$. The two-point contribution is given by 
\begin{align}\label{eqn:2ptPhiParSegment}
    \partial_{t_1}\partial_{t_2} L_{O_n}^{\text{2-pt}} 
    = - b_{n,2}^2 \epsilon^{2\Delta_n - 2} 
    \langle
    O_n(t_1 +  \epsilon){O_n}(t_2 -  \epsilon)
    \rangle_\oneD
    \;.
\end{align}
At finite coupling, {\it i.e.}~when ${\Delta_n>1}$, this integral is proportional to a positive power of $\epsilon$, and is therefore zero in the limit $\epsilon\to0$. It drops out from our derivation.
The three-point integral contribution is 
\begin{multline}\label{eqn:3ptSegment}
    \partial_{t_1}\partial_{t_2} L_{O_n}^{\text{3-pt}} 
    = 
    -b_{n,2}\epsilon^{\Delta_n - 1}\int_{t_1 + 2 \epsilon}^{t_2 - 2\epsilon}ds_2
    \bigg[
     \langle O_{\Phi_\perp^1}(t_1 +  \epsilon)O_{\Phi_\perp^1}(s_2)O_n(t_2 -  \epsilon)    \rangle_\oneD 
    \\ 
    +
    \langle O_{\Phi_\perp^1}(t_1 +  \epsilon)O_n(s_2)O_{\Phi_\perp^1}(t_2 -  \epsilon)    \rangle_\oneD
    +
    \langle O_n(t_1 +  \epsilon)O_{\Phi_\perp^1}(s_2)O_{\Phi_\perp^1}(t_2 -  \epsilon)    \rangle_\oneD
    \bigg]\; ,
\end{multline}
and is evaluated piece by piece in equation~\eqref{eqn:3ptIntegratedSegment}, we report the final result here:
\begin{align}\label{eqn:3ptIntegratedSegment0}
    \partial_{t_1}\partial_{t_2} L_{O_n}^{\text{3-pt}} 
    = 
    -(2 \mathbb{B})\,b_{n,2}\epsilon^{\Delta_n - 1}\frac{4\,C_n}{\Delta_n - 1} \,
    {\rm P}(t_1+ \epsilon,t_2- \epsilon)
    \;.
\end{align}
Finally, the last contribution is given by the integrated four-point function that reads 
\begin{align}\label{eqn:pink2}
    \partial_{t_1}\partial_{t_2} L^{\text{4-pt}}
    =-
    \int_{t_1 +3\epsilon}^{t_2 - 2\epsilon}\!\! ds_3 \int_{t_1 +2\epsilon}^{s_3 - \epsilon} \!\! ds_2
    \langle O_{\Phi_\perp^1}(t_1 +  \epsilon)O_{\Phi_\perp^1}(s_2)O_{\Phi_\perp^1}(s_3)O_{\Phi_\perp^1}(t_2 -  \epsilon) \rangle_\oneD
    \;.
\end{align}
The evaluation of this integral is explained in the next Appendix, in particular  see equation~\eqref{eqn:pink} which gives the relevant combination of $\partial_{t_1} \partial_{t_2} L^{\text{4-pt}} -\frac{1}{2} \partial_{t_1} \partial_{t_2} [ K^{\texttt{2-pt}} ]^2 $ appearing in our calculation.

At the end of these painstaking calculations, summing all together to reconstruct $$\partial_{t_2}\partial_{t_1} \left[B(t_1, t_2) - \frac{1}{2} A^2(t_1,t_2) \right],$$ all the divergences cancel. Indeed, the only divergent contributions appear in \eqref{eqn:3ptPhiSegmentEps} and \eqref{eqn:2ptSqMixed} that combined together give the following simple result
\beq
\partial_{t_2}\partial_{t_1} \bigg[ L_{\mathds{1}}^{\text{3-pt}}  - K_{\mathds{1}}^{\texttt{1-pt}}\times K^{\texttt{2-pt}}  \bigg] = -16 \, \mathbb{B} ^2 \, {\rm P}(t_1+ {\epsilon},t_2- {\epsilon})\;.
\eeq
Then, summing up all the remaining terms, we get
\beqa\label{eq:finalABresult}
&& \frac{\partial_{t_2}\partial_{t_1}\bigg[B(t_1,t_2) - \frac{A^2(t_1,t_2)}{2}\bigg] }{{\rm P}(t_1,t_2) } \\ &=&- (2 \mathbb{B} ) \sum_{\Delta_n>1} \;  b_{n,2}\;\frac{4\, C_{n} }{\Delta_n -1}  - (2 \mathbb{B} )^2 \, \left(-1+ \int_0^{\frac{1}{2}} dx \frac{{\delta} G(x) }{x^2} \log{\left(\frac{x^3}{1-x}\right)} \right)\, dx   .\nn
\eeqa
The previous expression can also be written in terms of the reduce correlator $f(x)$. Indeed, using \eqref{pt4} and integrating by parts, we obtain
 \beq\begin{split}
  &\frac{\partial_{t_2}\partial_{t_1}\bigg[B(t_1,t_2) - \frac{A^2(t_1,t_2)}{2}\bigg] }{{\rm P}(t_1,t_2) } =- 2 \mathbb{B} \! \sum_{\Delta_n>1}   b_{n,2}\frac{ 4\, C_n }{\Delta_n -1} - (2 \mathbb{B})^2 (2 \!-\! \mathbb{F} )\left( 1\! +\! \log 4 \right) \!+\! 4 \mathbb{B}^2  \\
&\qquad\qquad- (2 \mathbb{B} )^2 \,\left[ \int_{\delta_x}^{\frac{1}{2}} dx \frac{(2 x-3) ((x-1) x+1) \delta f(x)}{(x-1) x^3}\, dx + \frac{3}{2} (3 - \mathbb{F} ) \log(\delta_x) \right]\;,
\end{split}\eeq
where the integral in $x$ is finite in the limit $\delta_x\rightarrow 0^+$ since $\delta f(x)\sim \frac{3-\mathbb{F}}{2}x^2$ for $x\rightarrow 0$.

\section{Useful integrals}
\label{app:integrals}
\paragraph{Cross ratio on the circle.}In the following, to simplify some integrals over 4-point functions, it will be useful to recall the way the cross ratio is related to four points on the circle:
\begin{align}
\begin{split}
\label{eqn:Xcircledef}
X(s_1,s_2,s_3,s_4) &= \sqrt{\frac{{\rm P}(s_2,s_4){\rm P}(s_1,s_3)}{{\rm P}(s_3,s_4){\rm P}(s_1,s_2)}}\\
&= \frac{(e^{i\,s_1} - e^{i\,s_2})(e^{i\,s_3} - e^{i\,s_4})}{(e^{i\,s_1} - e^{i\,s_3})(e^{i\,s_2} - e^{i\,s_2})}\;.
\end{split}
\end{align}
\subsection{Integrals on the circle with two  insertions}\label{app:integrals1}
Here we collect and compute the integrals used in the arguments of section \ref{sec:firstargument}. 
\paragraph{Integrated 3-point function.}
The contribution of the 3-point functions in (\ref{eq:constraint8}) is proportional to the following one-dimensional integral
\begin{align}\label{eq:defI3}
\begin{split}
\mathcal{I}_{O_n} &=\int_0^{2 \pi} ds \; \mu_{1,2} \; {\rm P}(t_1,s)^{\frac{\Delta_n}{2}}\, {\rm P}(t_2,s)^{\frac{\Delta_n}{2}}\,{\rm P}(t_1,t_2)^{1-\frac{\Delta_n}{2}}\\
&=\left(\int_{t_1+\epsilon}^{t_2 - \epsilon}+ 
\int_{t_2+\epsilon}^{t_1 +2\pi - \epsilon}\right) ds \; {\rm P}(t_1,s)^{\frac{\Delta_n}{2}}\, {\rm P}(t_2,s)^{\frac{\Delta_n}{2}}\,{\rm P}(t_1,t_2)^{1-\frac{\Delta_n}{2}} 
\;,
\end{split}
\end{align}
where the integration measure $\mu_{1,2}$ enforcing the cutoff is defined in \eqref{munm}. 
This integral can be evaluated easily at leading order in the cutoff considering $\Delta_n>1$ a generic real number 
and it gives
\beq\label{eq:I3res}
\mathcal{I}_{O_n} = \frac{4 \,  {\epsilon}^{1 - \Delta_n}}{\Delta_n - 1}\; {\rm P}(t_1,t_2) + \mathcal{O}\left( {\epsilon}^{2-\Delta_n}\right) .
\eeq
While this result is divergent, it combines with the prefactor $\epsilon^{\Delta_n-1}$ in the action to produce the following finite contribution
\beq\label{eq:first3ptintegralsapp}
\mathcal{I}_{3-\text{pt}} = \sum_{\Delta_n>1} C_n \, (2 \mathbb{B} )\,  \epsilon^{\Delta_n-1} \mathcal{I}_{O_n}  \simeq (2 \mathbb{B} ) \,{\rm P}(t_1,t_2)\;\sum_{\Delta_n>1} b_{n,2} \frac{4\, C_n}{\Delta_n-1} \;.
\eeq

\paragraph{Integrated 4-point function. }
We also encountered the integral 
\begin{equation}\label{eqn:I4pt}
\mathcal{I}_{4-\text{pt}}\equiv \int_{0<s_1<s_2<2 \pi} \!\!\!\!\!\!\!ds_1 ds_2 \, 
\mu_{2,2}
\langle \Phi_{\perp}^M(t_1) \Phi_{\perp}^1(s_1)  \Phi_{\perp}^1(s_2) \Phi_{\perp}^M(t_2)  \rangle_\oneD ,
\end{equation}
with the integration measure defined in (\ref{mun}). 
Taking into account different orderings, ${\cal I}_{4-\text{pt}}$ can be rewritten as a sum of three terms
\begin{align}\label{eq:I4def}
\mathcal{I}_{4-\text{pt}} &=\; {\Big [ }  \int_{t_2<s_1<s_2<2 \pi+t_1} ds_1 ds_2\,
{\mu}_{2,2} 
\, G_1(t_1,t_2,s_1,s_2)\, {\rm P}(t_1,t_2) \, {\rm P}(s_2,s_1) \nn \\
&+\int_{t_1<s_1<t_2, \;t_2<s_2<2 \pi+t_1} ds_1 ds_2 \,
{\mu}_{2,2}
\,G_2(t_1,s_1,t_2,s_2)\, \, {\rm P}(t_1,s_1) \, {\rm P}(s_2,t_2)  \\
&+\int_{t_1<s_1<s_2<t_2 } ds_1 ds_2 \,
{\mu}_{2,2}
\, G_3(t_1,s_1,s_2,t_2)\,{\rm P}(t_1,s_1)\,{\rm P}(t_2,s_2) {\Big ]}\;(2 \mathbb{B} )^2 \;,\nn 
\end{align}
where $G_1$, $G_1$ and $G_3$ are defined in~\eqref{G1G2G3}.
It is convenient to split the $G$ functions as $G_i(x) = G_{i,\text{tree}}(x) + \delta G_i(x)$. Correspondingly we redefine \eqref{eq:I4def} as
\begin{equation}
\mathcal{I}_{4-\text{pt}}   \equiv (2\mathbb{B} )^2(\texttt{Tree} + \texttt{Loops})\;,
\end{equation}
where $\texttt{Tree}$ and $\texttt{Loops}$ are \eqref{eq:I4def} with the substitutions $G_i\rightarrow G_{i,\text{tree}}$ and  $G_i\rightarrow \delta G_{i}$ respectively.

Tree level contribution is the most singular, then we evaluate it separately. It can be easily computed using the values of $G_{i,\text{tree}}$ given in \eq{G1G2G3tree}. It boils down to elementary integrals and, for small cutoffs, it gives
\beqa
\texttt{Tree} &=& \, \left[ \frac{2 \pi}{\epsilon}  + \log\left( \epsilon^2 {\rm P}(t_1,t_2) \right) -6\right]\;{\rm P}(t_1,t_2) +{\rm O}(\epsilon) \;.
\eeqa
The remaining part containing $\delta G_i$ can be rewritten in terms of integrals over the cross ratio $x$. To do this, we change variables from $\{s_1, s_2\}$ to $\{s_1, x\}$
with 
\beq\label{eq:xdefapp2}
x \equiv X({t}_1 + \epsilon, s_1, s_2, {t}_2 - \epsilon)\;,
\eeq
where $X$ is defined in~\eqref{eqn:Xcircledef}. Explicitely, inverting the above relation we have
\begin{align}
    \label{eqn:sbs2}
    s_2(s_1,x) = 
    -i \log \left[-\frac{-x e^{i \left(s_1+t_1\right)}+e^{i
   \left(s_1+t_2\right)}+e^{i \left(t_1+t_2\right)} (x-1)}{e^{i s_1}
   (x-1)-e^{i t_2} x+e^{i t_1}}\right]\;.
\end{align}
Next, we need to work out  the range of integration in the new variables. 
\begin{figure}
    \centering
    \includegraphics[width=\columnwidth]{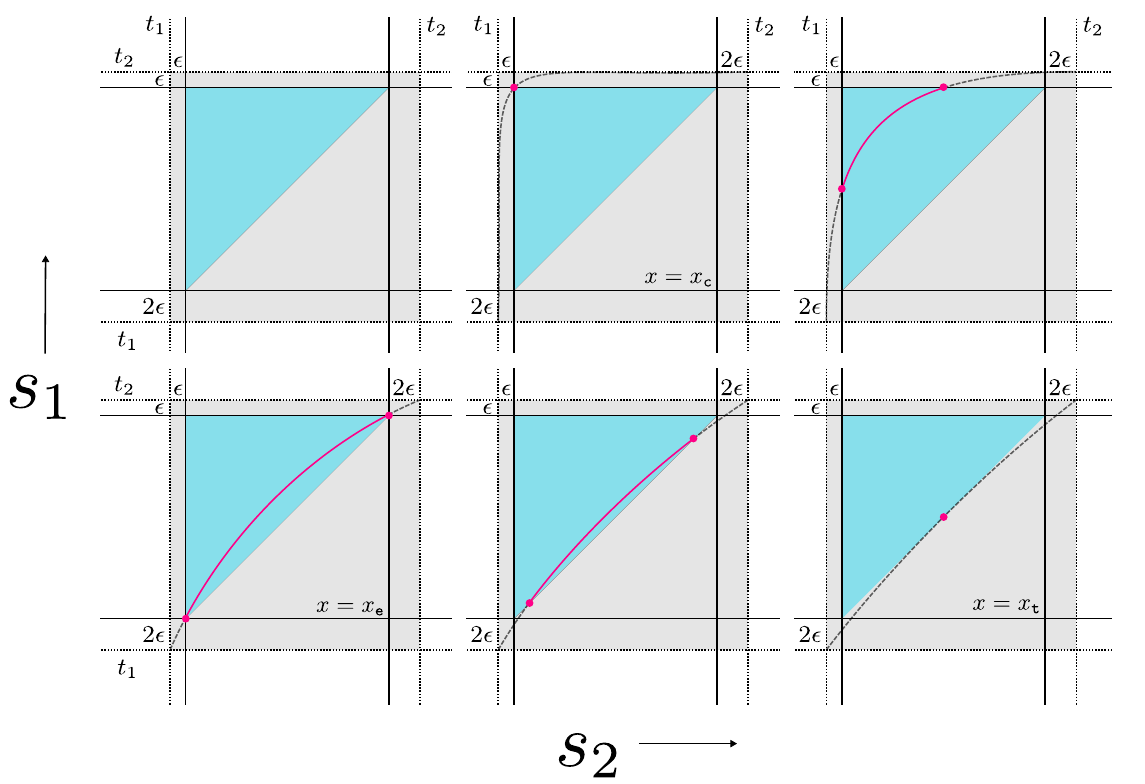}
    \caption{The integration range in the $(s_1,s_2)$ plane,~\eqref{eqn:ranges1s2}, is denoted by the shaded upper triangular region in blue. For a given fixed value of the cross ratio $x$, $(s_1, s_2)$ trace a curve ${\cal C}_x$, denoted by the pink arcs in the panels above. The intersection points of ${\cal C}_x$ with the triangular region~\eqref{eqn:ranges1s2} determine the limits of the $s_1$ integral for fixed $x$, and are denoted by pink circles.}
    \label{fig:ranges}
\end{figure}
The original integration domain for $s_1$ and $s_2$ is the triangular region given by
\begin{align}
    \label{eqn:ranges1s2}
    s_1\in [t_1 + \epsilon,t_2 - 2\epsilon] \;,\quad\text{and}\quad s_2\in [t_1 + 2 \epsilon,t_2 - \epsilon]\;, \quad \text{with}\quad s_2>s_1 + \epsilon\;,
\end{align}
see  figure~\ref{fig:ranges}. 
In the new variables $\{s_1,x\}$, this region~\eqref{eqn:ranges1s2} is charted in a rather non-trivial way. We will do first the $s_1$-integration, so we need the range in $s_1$ for fixed $x$. 
In figure~\ref{fig:ranges}, we depict the curve ${\cal C}_x$ in the $(s_1,s_2)$ plane given by $s_2 = s_2(x, s_1)$. The integration limits  for $s_1$ are the projections on the $s_1$-axis of the two points  where ${\cal C}_x$ enters and exits the triangular integration region~\eqref{eqn:ranges1s2}. 

 Depending on the values of $x$, $\mathcal{C}_x$ intersects the boundaries of the triangle on different sides (or lie completely outside of it). Accordingly, we need to split the $(x, s_1)$ integration in the following way ({{\it cf.}~figure~\ref{fig:ranges}}): 
\begin{itemize}
    \item {\tt Region 1:} Here,  $x_{\tt c} \leq x \leq x_{\tt e}$. The point $x = x_{\tt c}$ is the point such that  ${\cal C}_x$ intersects the upper-left corner of the triangle, $(s_1,s_2) = (t_1 + \epsilon,t_2-\epsilon)$, while  $x_{\tt e}$ is the point where ${\cal C}_x$ intersects the lower-left corner of the triangle $(s_1,s_2) = (t_1 + \epsilon, t_2 + 2\epsilon)$. In this range, ${\cal C}_x$ enters  the triangle through its left side and exits through the top side.
        \item {\tt Region 2:} Here, $x_{\tt e} \leq x \leq x_{\tt t}$, where $x_{\tt e}$ is defined as above and $x_{\tt t}$ is the point where ${\cal C}_x$ is tangent to the line $s_2 = s_1 + \epsilon$. In this range, ${\cal C}_x$ enters and exits the triangle through the diagonal side.
\end{itemize}
The values of $ x_{\tt t}$, $x_{\tt e}$, $x_{\tt c} $ are easy to obtain explicitly. We will need only their expansion for small $\epsilon $:
\beqa
  x_{\tt c} &=& \frac{\epsilon ^2}{2-2 \cos \left(t_1-t_2\right)}+O\left(\epsilon
   ^3\right)\sim 0
    \;,
\\
    x_{\tt e} &=& 
    \frac{1}{2}+\frac{\epsilon}{4}   \left(\cot \frac{t_1-t_2}{2}\right)+\frac{\epsilon ^2}{4 \cos
   \left(t_1-t_2\right)-4}+ {\rm O}\left(\epsilon ^3\right)
    \; \sim \frac{1}{2}, \\
       x_{\tt t} &=& 
    1+\epsilon \left( \cot \frac{t_1-t_2}{4}\right)
   +  \frac{\epsilon ^2}{2} \left( \cot
   ^2 \frac{t_1 - t_2}{4} \right)+{\rm O}\left(\epsilon
   ^3\right)\;\sim 1 .
\eeqa
The region of integration is thus split naturally in two. For $x_{\tt c} \leq x \leq x_{\tt e}$ (\texttt{Region 1}), we have $s \in [\check{s}_{\tt i}, \check{s}_{\tt f} ]$, where $\check{s}_{\tt i}$ and $\check{s}_{\tt f}$ are given by the intersection points of ${\cal C}_x$, with the lines $s_1=t_1 + \epsilon$ and $s_2 = t_2 - \epsilon$, respectively. These are given by
\begin{align}
    \check{s}_{\tt i} &= t_1 - \epsilon\;,\\
    \check{s}_{\tt f} &= t_2-\frac{\epsilon }{x}+ \epsilon^2\left(\frac{x-1}{2\, x^2}  \cot 
   \frac{t_1-t_2}{2} \right)+ {\rm O}\left(\epsilon ^3\right)\;.
\end{align}
For $x_{\tt e} \leq x \leq x_{\tt t}$ (\texttt{Region 2}), we have $s_1\in [\hat{s}_{\tt i}, \hat{s}_{\tt f}]$, with  these points defined by the intersections of ${\cal C}_x$, with the line $s_2=s_1 + \epsilon$. These are given by
\begin{align}
    \label{eqn:lim1s1}
    \hat{s}_{\tt i} = t_1+ \epsilon\left( \frac{x }{1-x} \right)-\epsilon^2 \left({\frac{x}{2\, (x-1)^2}} \cot \frac{t_1-t_2}{2}\right)
   +{\rm O}(\epsilon ^3)\;,\\
   \hat{s}_{\tt f} = t_2+\epsilon\left(\frac{1 }{x-1}\right)+
   \epsilon ^2
   \left(\frac{x}{2\,(x-1)^2}
   \cot \frac{t_1-t_2}{2}\right)
   +{\rm O}(\epsilon ^3)\;.
\end{align}
Putting all together, the original  integration  over the domain~\eqref{eqn:ranges1s2} rewrites as
\begin{align}
    \int_{t_1 + 2\epsilon}^{t_2 - \epsilon} 
    ds_2
    \int_{t_1 +  \epsilon}^{s_2 -\epsilon} ds_1
    = \left[
    \int_{x_{\tt e}}^{x_{\tt t}} dx 
    \int_{\hat{s}_{\tt i}}^{\hat{s}_{\tt f}} ds_1
    {\cal J}(s_1,x)
    \right]
    +
    \left[
    \int_{x_{\tt c}}^{x_{\tt e}} dx 
    \int_{\check{s}_{\tt i}}^{\check{s}_{\tt f}} ds_1
    {\cal J}(s_1,x)
    \right]
    \;,
\end{align}
where ${\cal J}(s_1,x)$ is the Jacobian of the transformation, given by 
\beqa
    {\cal J}(s_1,x) &=& \left| \frac{\partial s_2(s_1, x) }{\partial x} \right| \\&=&  -\frac{\sin \left(s_1-t_1\right)-\sin \left(s_1-t_2\right)+\sin
   \left(t_1-t_2\right)}{(x-1) \left(x \cos \left(s_1-t_2\right)-\cos
   \left(s_1-t_1\right)\right)+x \cos \left(t_1-t_2\right)-x^2+x-1} \;.\nn
\eeqa
Armed with this change of variables formula, we are now able to simplify the various four-point integrals in (\ref{eq:I4def}). 
Notice that the expressions for $x_{\tt c}$, $x_{\tt e}$ and $x_{\tt t}$ go to $0$, $1/2$ and $1$ respectively, as $\epsilon\to 0$. As we compute the various integrals that enter~\eqref{eq:I4def} in the small-$\epsilon$ limit, these values will naturally appear as limits on the $x$-integration.

Let us list the result for the three terms in  (\ref{eq:I4def}). Starting from the third line, after removing the tree-level part, we get the integral
$$
\int_{t_1 + 2\epsilon}^{t_2 - \epsilon} 
ds_2
\int_{t_1 +  \epsilon}^{s_2 -\epsilon} ds_1
\, \delta G_3(t_1,s_1,s_2,t_2)\; {\rm P}(t_1,s_1)\; {\rm P}(t_2,s_2).
$$
Reverting to the $(x, s_1)$ coordinates and doing the $s_1$ integration, the finite and divergent parts are
\begin{equation}\small\begin{aligned}
&\int_{t_1 + 2\epsilon}^{t_2 - \epsilon} 
ds_2
\int_{t_1 +  \epsilon}^{s_2 -\epsilon} ds_1
\, \delta G_3(t_1,s_1,s_2,t_2)\; {\rm P}(t_1,s_1)\; {\rm P}(t_2,s_2)\\
&={\rm P}(t_1,t_2)\left[\int_0^{\frac{1}{2}} dx \frac{\delta G_3(x)}{x^2} \;  \log\left(\frac{x}{ {\epsilon}^2\; {\rm P}(t_1,t_2)}\right) - \int_{\frac{1}{2}}^1 dx \frac{\delta G_3(x)}{x^2} \;\log \left(\frac{x \epsilon ^2
   {\rm P}(t_1,t_2)}{(x-1)^2}\right)\right]
   +{\rm O}(\epsilon)
   \;.
\label{eq:third}
\end{aligned}\normalsize\end{equation}
With the same method we evaluate the integral in the second line of (\ref{eq:I4def}), after subtracting the tree level contribution. This gives 
\begin{equation}\small\begin{aligned}
&\int_{t_1 + \epsilon}^{t_2-\epsilon} 
ds_1
\int_{t_2 + \epsilon}^{t_1 + 2 \pi - \epsilon} 
ds_2 
\,\delta G_2(t_1,s_1,t_2,s_2)\;{\rm P}(t_1,s_1) \; {\rm P}(t_2,s_2)\\
&={\rm P}(t_1,t_2)\left[\int_0^{\frac{1}{2}} dx \frac{\delta G_2(x)}{x^2} \;  \log\left(\frac{x {\epsilon}^2\;{\rm P}(t_1,t_2)}{1-x}\right)-\int_{\frac{1}{2}}^1 dx \frac{\delta G_2(x)}{x^2} \;\log\left(\frac{x {\epsilon}^2\; {\rm P}(t_1,t_2)}{1-x}\right)\right]
+{\rm O}(\epsilon)
\;,
\label{eq:second}
\end{aligned}\normalsize\end{equation}
while the integrals coming from the first line of (\ref{eq:I4def}) evaluate to 
\begin{equation}\small\begin{aligned}
&\int_{t_2 + 2\epsilon}^{t_1 + 2 \pi -\epsilon} ds_2 \int_{t_2 + \epsilon}^{s_2-\epsilon}  ds_1
\, \delta G_1(t_1,t_2,s_1,s_2)\; {\rm P}(t_1,t_2)\; {\rm P}(s_2,s_1)\\
&={\rm P}(t_1,t_2)\left[\int_0^{\frac{1}{2}} dx \frac{\delta G_1(x)}{x^2} \;  \log\left(\frac{x^2}{(1-x){\epsilon}^2\; {\rm P}(t_1,t_2)}\right) + \int_{\frac{1}{2}}^1 dx \frac{\delta G_1(x)}{x^2} \;\log\left(\frac{1 - x}{ {\epsilon}^2   {\rm P}(t_1,t_2)}\right) \right]+{\rm O}(\epsilon)\;.\label{eq:first}
\end{aligned}\normalsize\end{equation}
So in total, summing (\ref{eq:third})-(\ref{eq:first}), we have
\begin{align}
\texttt{Loops}
&=  \left[ -\log\left( \epsilon^2 {\rm P}(t_1,t_2) \right) \right]\, {\rm P}(t_1,t_2)\\
&\!\!\!\!\!+ \int_0^{\frac{1}{2} } \!\!dx \left[ \frac{\delta G_3(x) }{x^2} \log{x} + \frac{\delta G_1(x) }{x^2} \log\!{\left(\!\frac{x^2}{1-x}\!\right)} + \frac{\delta G_2(x) }{x^2} \log\!{\left(\!\frac{x}{1-x}\!\right)} \right]\! {\rm P}(t_1,t_2) +{\rm O}(\epsilon) \;,\nn
\end{align}
where we now used the crossing properties of the $G_i(x)$ amplitudes to map all the $x$ integrations to the interval $[0, \frac{1}{2} ]$. 
Altogether, therefore we find
\begin{align}\label{eq:I4result}
&\mathcal{I}_{4-\text{pt}}
=(2\mathbb{B} )^2\frac{2 \pi - 6 \epsilon }{\epsilon}{\rm P}(t_1,t_2)\\
&+ (2\mathbb{B} )^2\int_0^{\frac{1}{2} } \!\!dx \left[ \frac{\delta G_3(x) }{x^2} \log{x} + \frac{\delta G_1(x) }{x^2} \log\!{\left(\!\frac{x^2}{1-x}\!\right)} + \frac{\delta G_2(x) }{x^2} \log\!{\left(\!\frac{x}{1-x}\!\right)}\right]\!{\rm P}(t_1,t_2) +{\rm O}(\epsilon) \;.\nn
\end{align}
Using these explicit integrals, the constraint~(\ref{eq:constraint8}) in the main text becomes the the sum-rule~\eqref{eq:sumruleconstr}.

\subsection{Integrals on the segment}
We now discuss the explicit integrals entering the derivation in section \ref{sec:secondder}. 
\paragraph{Integrated 3-point function.}The first integral to compute is the 3-point contribution~\eqref{eqn:3ptSegment}, we display it below for convenience:
\begin{multline}\label{eqn:3ptSegment2}
    \partial_{t_1}\partial_{t_2} L_{O_n}^{\text{3-pt}} 
    = 
    -b_{n,2}\epsilon^{\Delta_n - 1}\int_{t_1 + 2 \epsilon}^{t_2 - 2\epsilon}ds_2
    \bigg[
     \langle O_{\Phi_\perp^1}(t_1 +  \epsilon)O_{\Phi_\perp^1}(s_2)O_n(t_2 -  \epsilon)    \rangle_\oneD 
    \\ 
    +
    \langle O_{\Phi_\perp^1}(t_1 +  \epsilon)O_n(s_2)O_{\Phi_\perp^1}(t_2 -  \epsilon)    \rangle_\oneD
    +
    \langle O_n(t_1 +  \epsilon)O_{\Phi_\perp^1}(s_2)O_{\Phi_\perp^1}(t_2 -  \epsilon)    \rangle_\oneD
    \bigg]\;.
\end{multline}
It is a sum of three terms. 
The individual integrals are computed below.
The first term of~\eqref{eqn:3ptSegment2}, gives
\begin{multline}
    \epsilon^{\Delta_n-1}
    \int_{t_1+  2\epsilon}^{t_2 -  2\epsilon} d\,s_2\,
    {\rm P}(t_1 +  \epsilon,s_2)^{1-\frac{\Delta_1}{2}}\,
    {\rm P}(s_2,t_2 -   \epsilon)^{\frac{\Delta_1}{2}}\,
    {\rm P}(t_1 +   \epsilon,t_2 -  \epsilon)^{\frac{\Delta_1}{2}}
    \\
    = \frac{1}{\Delta_n - 1}{\rm P}(t_1 +  \epsilon, t_2 -  \epsilon) 
    + \mathcal{O}\left(\epsilon^{2\Delta_n - 2}\right)\;.
\end{multline}
The second term of~\eqref{eqn:3ptSegment2}, gives
\begin{multline}
    \epsilon^{\Delta_n-1}
    \int_{t_1+  2\epsilon}^{t_2 -  2\epsilon} d\,s_2\,
    {\rm P}(t_1 +  \epsilon,s_2)^{\frac{\Delta_1}{2}}\,
    {\rm P}(s_2,t_2 -   \epsilon)^{\frac{\Delta_1}{2}}\,
    {\rm P}(t_1 +   \epsilon,t_2 -  \epsilon)^{1-\frac{\Delta_1}{2}}
    \\
    = \frac{2}{\Delta_n - 1}{\rm P}(t_1 +  \epsilon, t_2 -  \epsilon) 
    + \mathcal{O}\left(\epsilon^{2\Delta_n - 2}\right)\;,
\end{multline}
while the third term gives
\begin{multline}
    \epsilon^{\Delta_n-1}
    \int_{t_1+  2\epsilon}^{t_2 -  2\epsilon} d\,s_2\,
   {\rm P}(t_1+ \epsilon,s_2)^{\frac{\Delta_1}{2}}\,
    {\rm P}(s_2,t_2- \epsilon)^{1-\frac{\Delta_1}{2}}\,
    {\rm P}(t_1+ \epsilon,t_2- \epsilon)^{\frac{\Delta_1}{2}}
    \\
    = \frac{1}{\Delta_n - 1}{\rm P}(t_1+ \epsilon,t_2- \epsilon)
    + \mathcal{O}\left(\epsilon^{2\Delta_n - 2}\right)\;.
\end{multline}
For their combination, we get
\begin{align}\label{eqn:3ptIntegratedSegment}
    \partial_{t_1}\partial_{t_2} L_{O_n}^{\text{3-pt}} 
    = 
    -(2 \mathbb{B})\,b_{n,2}\epsilon^{\Delta_n - 1}\frac{4\,C_n}{\Delta_n - 1} \,
    {\rm P}(t_1+ \epsilon,t_2- \epsilon)
    \;.
\end{align}

\begin{figure}
    \centering
    \includegraphics[width=\columnwidth]{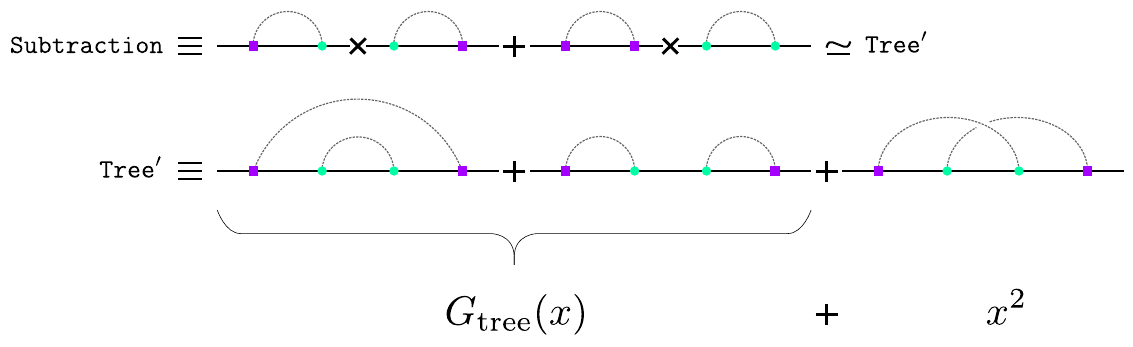}
    \caption{Here we illustrate schematically some terms appearing in the calculation (with purple squares $\simeq$ fixed variables, and green circles $\simeq$ integrated variables). The term \texttt{Tree}$^\prime$ is given by the tree-level 4-point function, integrated in the two middle variables, plus an extra term given by two propagators crossed in a non-planar fashion. Superficially, these terms have the same form as the terms of the \texttt{Subtraction}, which are also given in the first line in terms of products of integrated 2-point functions integrated in some of the variables. The only mismatch comes from the range of the integrated variables, cf. (\ref{eq:Treeprime}) vs (\ref{eq:subtractioneq}). The tiny $\epsilon$-dependent mismatch, combined with the divergence of the integrals, produces a finite result  in our calculation, cf. (\ref{eq:tinymismatch}).  
    \label{fig:treeprime}
    }
\end{figure}

\paragraph{Integrated 4-point function.}
Our starting point is the integral~\eqref{eqn:pink2},
which we repeat below for convenience
\beq
\partial_{t_1} \partial_{t_2} L^{\text{4-pt}} \equiv  -
    \int_{t_1 + 3\epsilon}^{t_2 - 2\epsilon} ds_3 \int_{t_1 +2\epsilon}^{s_3 - \epsilon}  ds_2
    \langle O_{\Phi_\perp^1}(t_1 +  \epsilon)O_{\Phi_\perp^1}(s_2)O_{\Phi_\perp^1}(s_3)O_{\Phi_\perp^1}(t_2 -  \epsilon) \rangle_\oneD \nn\;.
\eeq
More explicitly, we have 
\beq
\partial_{t_1} \partial_{t_2} L^{\text{4-pt}} \equiv  - (2 \mathbb{B} )^2\;  \int_{t_1 + 3\epsilon}^{t_2 - 2\epsilon} ds_3 \int_{t_1 +2\epsilon}^{s_3 - \epsilon}  ds_2 \; {G(x) }\;{\rm P}({t}_1 +\epsilon , s_3) \, {\rm P}({t}_2 - \epsilon , s_2)\;,
\eeq
Again it is convenient to separate the contribution of the tree level part (which is divergent). Moreover it is convenient on top of the tree level part to also include and subtract an extra pice. This piece corresponds to the  non-planar diagram shown last in the second line in Figure \ref{fig:treeprime}. 
Namely, we define
\beqa
 \partial_{t_1} \partial_{t_2} L^{\text{4-pt}}  \equiv (2 \mathbb{B} )^2 \times \left( \texttt{Tree}' + \texttt{Loops}' \right),
\eeqa
where
\beq
\texttt{Loops}' \equiv -\int_{t_1 +3 \epsilon}^{t_2 -2 \epsilon} ds_3 \int_{t_1 + 2\epsilon}^{s_3 - \epsilon}  ds_2 \; \left({\delta G(x) -x^2  }\right)\;{\rm P}(t_1 + \epsilon , s_2) \, {\rm P}(t_2 - \epsilon , s_3) \;,
\eeq
and 
\beq
\texttt{Tree}' \equiv -\int_{t_1 + 3 \epsilon}^{t_2 -2\epsilon} ds_3 \int_{t_1 + 2\epsilon}^{s_3 - \epsilon}  ds_2 \; \left({G_{\text{tree}}(x) + x^2 } \right)\;{\rm P}(t_1 + \epsilon , s_2) \, {\rm P}(t_2 - \epsilon , s_3) \;. 
\eeq
The extra $x^2$ term is particularly convenient to recombine $\texttt{Tree}'$ with another contribution, as we will see shortly. 

The most complicated part can be computed by the same method illustrated in the previous section, i.e. we trade one of the integration variables for an integral over the cross ratio $x$ defined by (\ref{eq:xdefapp2}). This results in \begin{align}
    \texttt{Loops}' &= -{\rm P}(t_1,t_2)\,\left( \int_0^{\frac{1}{2}} dx\, \frac{\delta G(x) - x^2 }{x^2} \log{\left(\frac{x^3}{1-x}\right)} \right) + \mathcal{O}(\epsilon)\nn\\
    &= -{\rm P}(t_1,t_2)\,\left( - \int_0^{\frac{1}{2}} dx\,\log{\left(\frac{x^3}{1-x} \right)}+ \int_0^{\frac{1}{2}} dx\, \frac{\delta G(x)}{x^2}\log{\left(\frac{x^3}{1-x} \right) } \right) + \mathcal{O}(\epsilon)\nn\\
    &= -{\rm P}(t_1,t_2)\,\left( 1 + \log 4 + \int_0^{\frac{1}{2}} dx\, \frac{\delta G(x)}{x^2} \log{\left(\frac{x^3}{1-x} \right)} \right) + \mathcal{O}(\epsilon)\;.
\end{align}
The tree level part is defined explicitly as
\begin{multline}\label{eq:Treeprime}
\texttt{Tree}' \equiv 
-\int_{t_1 + 3\epsilon}^{t_2 -2\epsilon} ds_3 \int_{t_1 +2\epsilon}^{s_3 - \epsilon}  ds_2 \; \left[ {\rm P}(t_1 + \epsilon,s_1)\,{\rm P}(t_2 - \epsilon,s_2) \right.\\\left.+  {\rm P}(t_1 + \epsilon,{{t}}_2-\epsilon)\,{\rm P}(s_1,s_2) +  {\rm P}(t_1 + \epsilon,s_2)\,{\rm P}(t_2 - \epsilon,s_1) \right] \;.
\end{multline}
This contribution can be evaluated explicitly with \texttt{Mathematica} and then expanded in the cutoffs. It is a complicated and divergent expression, however all divergences are cancelled by other terms coming from the expansion of $-\frac{1}{2} \partial_{t_1} \partial_{t_2} A^2(t_1, t_2)$ defined in (\ref{eq:Ksq}). What will be relevant for us is the combination:
\beq
\texttt{Tree}' -\underbrace{\frac{1}{2} \partial_{t_1} \partial_{t_2} \left[  \int_{t_1 +  \epsilon}^{t_2-2\epsilon} ds_2
    \int_{s_2 + \epsilon}^{t_2 -  \epsilon}  ds_3 \,{\rm P}(s_2,s_3) \right]^2 }_{\equiv \texttt{Subtraction}} ,
\eeq 
where the second term comes from $-\frac{1}{2} \partial_{t_1} \partial_{t_2} [ K^{\texttt{2-pt}} ]^2 \equiv - (2 \mathbb{B} )^2 \,(\texttt{Subtraction})$ in (\ref{eq:Ksq}),  and can be rewritten as
\beqa\label{eq:subtractioneq}
&&\texttt{Subtraction} = -{\rm P}(t_{1}+ {\epsilon},t_{2}- {\epsilon} ) \; \int_{t_1 +  \epsilon}^{t_2-2\epsilon} ds_2
    \int_{s_2 + \epsilon}^{t_2 -  \epsilon}  ds_3 \,{\rm P}(s_2,s_3) \\
    &&- \left( \int_{t_1 +2\epsilon < s_2<t_2- {\epsilon}}  {ds_2}\,{\rm P}(t_1 + \epsilon ,s_2)  \right)\times \left( \int_{t_1 + {\epsilon} < s_3<t_2-2\epsilon}   {d s_3}\,{\rm P}(t_2 - \epsilon , s_3) \right). \nn
\eeqa
As can be seen most easily graphically, see Figure \ref{fig:treeprime},  the combination $\texttt{Tree}' - \texttt{Subtraction}$ almost perfectly cancels. Notice however that the integration regions are slightly different, by an infinitesimal amount. This combines with the divergences of the integrals to give a finite result for the difference. Evaluating the integrals explicitly, we can compute  
\beq\label{eq:tinymismatch}
  \texttt{Tree}'-\texttt{Subtraction}   = {\rm P}(t_1,t_2) \; \left(6 + \log{4} \right) + \mathcal{O}(\epsilon) .
\eeq
All in all we have a finite result
\beqa
&&\partial_{t_1} \partial_{t_2} L^{\text{4-pt}} -\frac{1}{2} \partial_{t_1} \partial_{t_2} [ K^{\texttt{2-pt}} ]^2 = (2 \mathbb{B} )^2 \;\left( \texttt{Tree}'+\texttt{Loops}'-\texttt{Subtraction}  \right)\nn \\ 
&=& -(2 \mathbb{B} )^2 \, {\rm P}(t_1,t_2)\,\left( -5 + \int_0^{\frac{1}{2}} dx \frac{\delta G(x)}{x^2} \log{\left(\frac{x^3}{1-x}\right)} \right)\;.
\label{eqn:pink}
\eeqa

\section{Normalisation of $\Phi_{||}$: details}\label{app:appD}

\paragraph{Weak coupling.}
The anomalous term in \eqref{eq:nontribs} arise from the divergence of the superconformal block of the long-multiplet at $x\sim 0$ for $\Delta_1=1$. The divergent contribution is given by the following integral
\beq\label{III}
\int_{\delta_x}^{1/2} \frac{x-2}{1-\Delta_1 } x^{\Delta_1 -2}dx\;.
\eeq
At finite coupling $\Delta_1>1$, the integral is convergent and it gives
\beq
I_{\text{finite}}=\frac{2^{-\Delta_1 } (3 \Delta_1 +1)}{(\Delta_1 -1)^2 \Delta_1 }\;.
\eeq
At weak coupling $\Delta_1=1+\gamma_1$, we expand first for small $\gamma_1$ and then integrate. Choosing
the prescription in which $\log \delta_x\rightarrow 0$, we can resum order by order in $\gamma_1$ obtaining
\begin{align}
I_{\text{weak}}&=\frac{1}{2(\Delta_1\!-\!1)}+\sum_{k=1}^\infty\biggl[
\frac{(1-\Delta_1)^{k-1}}{2} \left(\!1+ \frac{4\log^k 2}{(1-\Delta_1)k!}\right)\!+\frac{\log ^k 2 }{2 k!}\!\!\sum_{n=k-1}^\infty\! (1-\Delta_1)^n\biggl]\nonumber\\
&=\frac{2^{-\Delta_1 } \left(1-\left(2^{\Delta_1 +1}-3\right) \Delta_1 \right)}{(\Delta_1 -1)^2
   \Delta_1 }\;.
\end{align}
The anomaly is proportional to the discrepancy of the two previous results
\beq
\delta I=I_{\text{finite}}-I_{\text{weak}}=
\frac{2}{(\Delta_1-1)^2}\;.
\eeq
Since the integral \eqref{III} comes from the OPE, it is multiplied by $C_1^2$ leading to the following identity
\beq\label{fint1}
 \left. \int_{\delta_x}^{1/2} \!
 \frac{(x-2)\delta f(x) }{x^3}\;dx   \right|_{ \text{small }g } \!\!\!\!\sim    \underbrace{ \int_{\delta_x}^{\frac{1}{2}}dx \frac{ (x-2)\sum_{\ell=1}^{M} g^{2 \ell} f_{\text{weak}}^{(\ell)}(x)}{x^3} }_{\texttt{regularised, }\log(\delta_x)\rightarrow 0} + \underbrace{\left[\frac{2 \;C_1^2}{(\Delta _1-1)^2}\right] }_{\texttt{``anomaly''}} + \mathcal{O}(g^{2 M + 2}) ,
\eeq
where $f_{\text{weak}}^{(\ell)}(x)$ is the weak coupling reduced correlator at $\ell$ loops given in  \cite{Cavaglia:2022qpg} together with the perturbative expansions of $C_1^2$ and $\Delta_1$. 

Using equation \eqref{fint1} together with \eqref{eq:nontribs}, we can compute the Wilson coefficient $b_{1,2}$ up to order $g^7$, obtaining the following expansion
\begin{align}
b_{1,2}&=\frac{g}{\sqrt{2}}-\frac{g^3\left(\pi ^2-6\right) }{3 \sqrt{2}}+\frac{g^5 \left(-108 (\zeta_3+3)+12 \pi ^2+5 \pi ^4+576 \log
  2\right)}{18 \sqrt{2}}\nonumber\\
   &+\frac{g^7}{270
   \sqrt{2}} \biggl[60 \pi ^2
   \left(132 \zeta_3+399+8 \log ^3 2-8  (36+3 \log 2)\log 2\right)-71 \pi ^6-138915 \zeta_5\nonumber\\
   &+6 \pi ^4 (91-188 \log 2)-72 \biggl(-15 \left(32 \text{Li}_4\left(\tfrac{1}{2}\right)-16
   \text{Li}_5\left(\tfrac{1}{2}\right)+16 S_{3,2}\left(\tfrac{1}{2}\right)+33 \zeta_3-271\right)\nonumber\\
   &+60 \log 2 \left(4
   \text{Li}_4\left(\tfrac{1}{2}\right)+\zeta_3 (2 \log 2-7)-4\right)+ (6 \log 2-20)\log ^4 2\biggl)\biggl]+\,\mathcal{O}\left(g^9\right)\;,
\end{align}
where $\text{Li}_n(x)$ are polylogarithms and $S_{n,m}(x)$ are Nielsen generalised polylogarithms (or hyperlogarithm).

\paragraph{Strong coupling.}
At strong coupling we obtain the following expansion
\begin{align}
&b_{1,2}=\frac{ g (1\!+\!\log 2)}{\sqrt{2/5}\,\pi }-\frac{ 149\!-\!8 \pi ^2\!+\!17 \log 2}{96\,\sqrt{2/5}\, \pi
   ^2}+\frac{ 5616 \zeta_3\!+\!21003\!-\!1232 \pi ^2\!+\!45 \log 2 (256 \log 2\!-\!129)}{18432\,\sqrt{2/5}\, \pi ^3
   g}\nonumber\\
   &+\frac{477631
   \log 2\!-\!864 \zeta_3 (515\!+\!76 \log 2)\!-\!1039805\!+\!65496 \pi ^2\!+\!2304 \log ^2 2 (216 \log 2\!-\!673)}{1769472 \,\sqrt{2/5}\,\pi ^4 g^2}\nonumber\\
   &+\frac{1}{679477248
   \sqrt{10} \pi ^5 g^3}\biggl[5 \biggl(53747712 \text{Li}_5\left(\tfrac{1}{2}\right)-71663616
   S_{3,2}\left(\tfrac{1}{2}\right)+183253536 \zeta_3+8024832 \zeta_5\nonumber\\
   &+223558211-96292333 \log 2\biggl)+32 \biggl(108 \log 2
   (77760 \text{Li}_4\left(\tfrac{1}{2}\right)+204295 \zeta_3+4 \log 2 (14850 \zeta_3\nonumber\\
   &+\!175355\!+\!16 \log 2 (\log 2 (2890\!+\!27
   \log 2)\!-\!10480)))\!-\!5 \pi ^2 \left(324864 \zeta_3\!+\!420965+46656 \log ^3 2\right)\nonumber\\
   &-15552 \pi ^4 \log 2\biggl)\biggl]+\,\mathcal{O}\left(\frac{1}{g^4}\right)\;.
\end{align}

\bibliographystyle{JHEP.bst}
\bibliography{references}

\end{document}